\documentclass[letter, longauth]{aa}

\relax
\usepackage{graphicx}
\usepackage{txfonts}
\usepackage{pdflscape}
\usepackage{color}

\usepackage{amstext}
\usepackage[colorlinks=true,linkcolor=black,citecolor=blue]{hyperref}%

\begin{document} 

\title{A high angular resolution view of the PAH emission in Seyfert galaxies using JWST/MRS data}

   \author{I. Garc\'ia-Bernete\inst{1}\fnmsep\thanks{E-mail: igbernete@gmail.com}, D. Rigopoulou\inst{1}, A. Alonso-Herrero\inst{2}, F.\,R. Donnan\inst{1}, P.\,F. Roche\inst{1}, M. Pereira-Santaella\inst{3,4}, A. Labiano\inst{5,3}, L. Peralta de Arriba\inst{2}, T. Izumi\inst{6}, C. Ramos Almeida\inst{7,8}, T. Shimizu\inst{9}, S. H\"onig\inst{10}, S. Garc\'ia-Burillo\inst{4}, D.\,J. Rosario\inst{11},  M.\,J. Ward\inst{12}, E. Bellocchi\inst{13,14}, E.\,K.\,S. Hicks\inst{15}, L. Fuller\inst{16} and C. Packham\inst{16,17}.}

   \institute{$^1$Department of Physics, University of Oxford, Keble Road, Oxford OX1 3RH, UK \\
   $^2$ Centro de Astrobiolog\'ia (CAB), CSIC-INTA, Camino Bajo del Castillo s/n, E-28692, Villanueva de la Ca\~nada, Madrid, Spain\\
   $^3$Centro de Astrobiolog\'ia, CSIC-INTA, Ctra de Torrej\'on a Ajalvir, km\,4, E-28850, Torrej\'on de Ardoz, Madrid, Spain\\
   $^4$Observatorio Astron\'omico Nacional (OAN-IGN)-Observatorio de Madrid, Alfonso XII, 3, 28014, Madrid, Spain\\
   $^5$Telespazio UK for the European Space Agency (ESA), ESAC, Camino Bajo del Castillo s/n, 28692 Villanueva de la Ca\~nada, Spain\\
   $^6$National Astronomical Observatory of Japan, 2-21-1 Osawa, Mitaka, Tokyo 181-8588, Japan\\
   $^7$Instituto de Astrof\'isica de Canarias, Calle V\'ia L\'actea, s/n, E-38205 La Laguna, Tenerife, Spain\\
   $^8$Departamento de Astrof\'isica, Universidad de La Laguna, E-38206 La Laguna, Tenerife, Spain\\
   $^9$Max-Planck-Institut fur extraterrestrische Physik, Postfach 1312, D-85741 Garching, Germany\\
   $^{10}$Department of Physics \& Astronomy, University of Southampton, Hampshire, SO17 1BJ, Southampton, UK\\
   $^{11}$ School of Mathematics, Statistics and Physics, Newcastle University, Herschel Building, Newcastle upon Tyne, NE1 7RU, UK\\
   $^{12}$Centre for Extragalactic Astronomy, Durham University, South Road, Durham DH1 3LE, UK\\
   $^{13}$ Departamento de F\'isica de la Tierra y Astrof\'isica, Fac. de CC F\'isicas, Universidad Complutense de Madrid, 28040, Madrid, Spain
   $^{14}$Instituto de F\'isica de Part\'culas y del Cosmos IPARCOS, Fac. CC F\'isicas, Universidad Complutense de Madrid, 28040 Madrid, Spain
   $^{15}$Department of Physics \& Astronomy, University of Alaska Anchorage, AK 99508-4664, USA\\
   $^{16}$The University of Texas at San Antonio, One UTSA Circle, San Antonio, TX 78249, USA\\
   $^{17}$National Astronomical Observatory of Japan, 2-21-1 Osawa, Mitaka, Tokyo 181-8588, Japan}

\titlerunning{A high angular resolution view of the PAH emission in Seyfert galaxies using JWST/MRS data}
\authorrunning{Garc\'ia-Bernete et al.}

   \date{}

  \abstract
   {Polycyclic aromatic hydrocarbons (PAHs) are carbon-based molecules that are ubiquitous in a variety of astrophysical objects and environments. In this work we use JWST/MIRI MRS spectroscopy of three Seyferts to compare their nuclear PAH emission with that of star-forming (SF) regions. This study represents the first of its kind to use sub-arcsecond angular resolution data of local luminous Seyferts (L$_{\rm bol}>10^{44.46}$\,erg/s) with a wide wavelength coverage (4.9-28.1 $\mu$m). We present an analysis of their nuclear PAH properties by comparing the observed ratios with PAH diagnostic model grids derived from theoretical spectra. Our results show that a suite of PAH features is present in the innermost parts of luminous Seyfert galaxies ($\sim$0.45\arcsec at 12\,$\mu$m; in the inner $\sim$142-245\,pc). We find that the nuclear regions of active galactic nuclei (AGN) lie at different positions of the PAH diagnostic diagrams, whereas the SF regions are concentrated around the average values of SF galaxies. In particular, we find that the nuclear PAH emission mainly originates in neutral PAHs. In contrast, PAH emission originating in the SF regions favours ionised PAH grains. The observed PAH ratios in the nuclear region of the AGN-dominated galaxy NGC\,6552 indicate the presence of larger PAH molecules compared with those of the SF regions. Therefore, our results provide evidence that AGN have a significant impact on the ionisation state (and probably the size) of the PAH grains on scales of $\sim$142-245\,pc.}

      
   \maketitle


\section{Introduction}
Polycyclic aromatic hydrocarbons (PAHs) are carbon-based molecules that typically include one or more carbon rings.
These molecules absorb a significant fraction of  UV/optical photons from (mainly) young stars (e.g. \citealt{Peeters04}), resulting in their excitation. The excited PAH molecules produce infrared (IR) features (the brightest bands are 3.3, 6.2, 7.7, 8.6, 11.3, 12.7, and 17.0~$\mu$m; e.g. \citealt{Tielens08}) through vibrational relaxation (e.g. \citealt{Draine07}). Thus, PAH features are considered excellent tracers of the star formation activity in star-forming (SF) galaxies (e.g. \citealt{Rigopoulou99,Peeters04}), but also in active galactic nuclei (AGN; e.g. \citealt{Diamond12,Esquej14}).

The relative variations between PAH features indicate different physical conditions (see e.g. \citealt{Li20} for a review). This makes PAH features a powerful tool for characterising the interstellar medium (ISM) in different astrophysical objects and environments. However,  our understanding of the effect of the hardness of the radiation field on these molecules is limited. Indeed, AGN-dominated systems generally show PAH emission with lower equivalent widths compared to those observed in SF galaxies. Therefore, it has been proposed that PAH features appear to be diluted by the strong AGN continuum (e.g. \citealt{Herrero14,Almeida14,Bernete15}) or that PAH molecules are destroyed by the hard radiation field of the AGN (e.g. \citealt{Roche91,Voit92,Siebenmorgen04}). In addition, it is possible that the AGN UV emission could contribute to the PAH excitation within $\sim$100\,pc of AGN \citep{Jensen17}.

More recently, \citet{Bernete22} found that the PAH molecules responsible for the 11.3\,$\mu$m PAH emission band are more resilient in the hard environments often present in AGN. In particular, the authors found larger 11.3/7.7\,$\mu$m and 11.3/6.2\,$\mu$m PAH ratios in AGN-dominated systems compared to SF galaxies,
indicating a larger fraction of neutral PAH molecules
(as noted by \citealt{Smith07a} using a sample of relatively weak AGN). However, these studies were limited by the spatial resolution ($\sim$4$\arcsec$) and the low spectral resolution (R$\sim$60--130) of \emph{Spitzer}/InfraRed Spectrograph (IRS). Previous sub-arcsecond angular resolution {{N-band}} ($\sim$8--13\,$\mu$m) ground-based spectroscopic studies investigated the 11.3\,$\mu$m PAH feature in the nuclear and circumnuclear regions of AGN (e.g. \citealt{Hoenig10,gonzalez-martin13,Herrero14,Herrero16,Almeida14,Esquej14,Bernete15,Jensen17,Esparza-Arredondo18}). However, these works were unable to provide definitive details regarding the effect of the AGN on the PAH molecules due to limited wavelength coverage and sensitivity. The changes in the PAH properties due to the presence of the AGN might be more prominent in their innermost regions of galaxies. Therefore, the unprecedented combination of high angular and spectral resolution (R$\sim$1500-3500) in the entire mid-IR range (4.9-28.1 $\mu$m) afforded  by the \emph{James Webb} Space Telescope (JWST)/Mid-Infrared
Instrument (MIRI; \citealt{Rieke15, Wells15, Wright15}) is key to investigating PAH properties. In this Letter we report on the first investigation of PAH emission in the nuclear regions of three luminous Seyfert (Sy) galaxies and compare them with emission from SF regions using JWST/MIRI Medium\ Resolution Spectrograph (MRS) data. This enables us, for the first time, to characterise the PAH properties of 
local luminous Sy galaxies (log (L$_{\rm bol})>44.46$\,erg/s)\footnote{Bolometric luminosities are obtained from the 14-195\,keV X-ray intrinsic luminosities by multiplying by a factor of 7.42 as in \citet{Bernete19}.}
at sub-arcsecond scales ($\sim$0.45\arcsec, $\sim$142-245\,pc). 

\section{Targets and observations}
\label{sample}
The present study employs all the MIRI/MRS observations of Sy galaxies that are currently publicly available in the JWST archive\footnote{\textcolor{blue}{https://mast.stsci.edu/portal/Mashup/Clients/Mast/Portal.html}}: NGC\,6552 (MIRI Commissioning Observations: Program ID 1039, PI D. Dicken), NGC\,7319 (Early Release Observations: Program ID 2732, PI K. M. Pontoppidan; \citealt{Pontoppidan22}), and NGC\,7469 (Early Release Science: Program ID 1328, PI L. Armus).

NGC\,6552 is a Sy2 galaxy with confirmed polarised (hidden) broad line region \citep{Tran01}. NGC\,7319 harbours a type 2 AGN with no evidence of starburst (SB) activity \citep{Sulentic01}. Finally, NGC\,7469 is a barred Sy1 galaxy (\citealt{Osterbrock93}) that hosts a nuclear SB (\citealt{Davies07}; r$\sim$50-65\,pc) and is surrounded by a 1.6~kpc SB ring (e.g. \citealt{Glass98,Diaz-Santos07}). The main properties of the three galaxies are shown in Table \ref{table_prop}. 

We retrieved MIRI/MRS data from the JWST archive. Then, we processed them using the JWST calibration pipeline (version 1.6.2) to produce the 12 sub-channel cubes with the default spatial and spectral sampling (see Appendix \ref{reduction} for a full description of the data reduction). For NGC\,7469, we also downloaded the fully reduced MIRI F770W observations from the JWST archive in order to produce the 7.7 $\mu$m PAH emission map of the SF ring.

\begin{table}[ht]
\centering
\begin{tabular}{lcccc}
\hline
Name & Spectral   &     D$_{\rm L}$ & log L$_{\rm 14-195\,keV}$ & log N$_{\rm H}^{\rm X-ray}$\\
 & type   &     (Mpc) &(erg~s$^{-1}$)& (cm$^{-2}$) \\
\hline
NGC\,6552 & Sy2& 118.0 & 43.65 &24.02\\
NGC\,7319 & Sy2& 96.1 & 43.76 &23.82\\
NGC\,7469 & Sy1.5& 67.2& 43.59 &20.53\\
\hline
\end{tabular}                                           
\caption{Main properties of the Sy galaxies used in this work. The spectral types were taken from \citet{Veron06}. The luminosity distance and spatial scale were calculated using a cosmology with H$_0$=70 km~s$^{-1}$~Mpc$^{-1}$, $\Omega_m$=0.3, and $\Omega_{\Lambda}$=0.7. The intrinsic L$_{\rm 14-195\,keV}$ and N$_{\rm H}^{\rm X-ray}$ were taken from \citet{Ricci17}.}
\label{table_prop}
\end{table}

\section{Region selection and spectral analysis}
\label{fitting}
\subsection{Nuclear regions of the Sy galaxies}

\begin{figure}
\centering
\par{
\includegraphics[width=9.3cm, clip, trim=100 10 10 10]{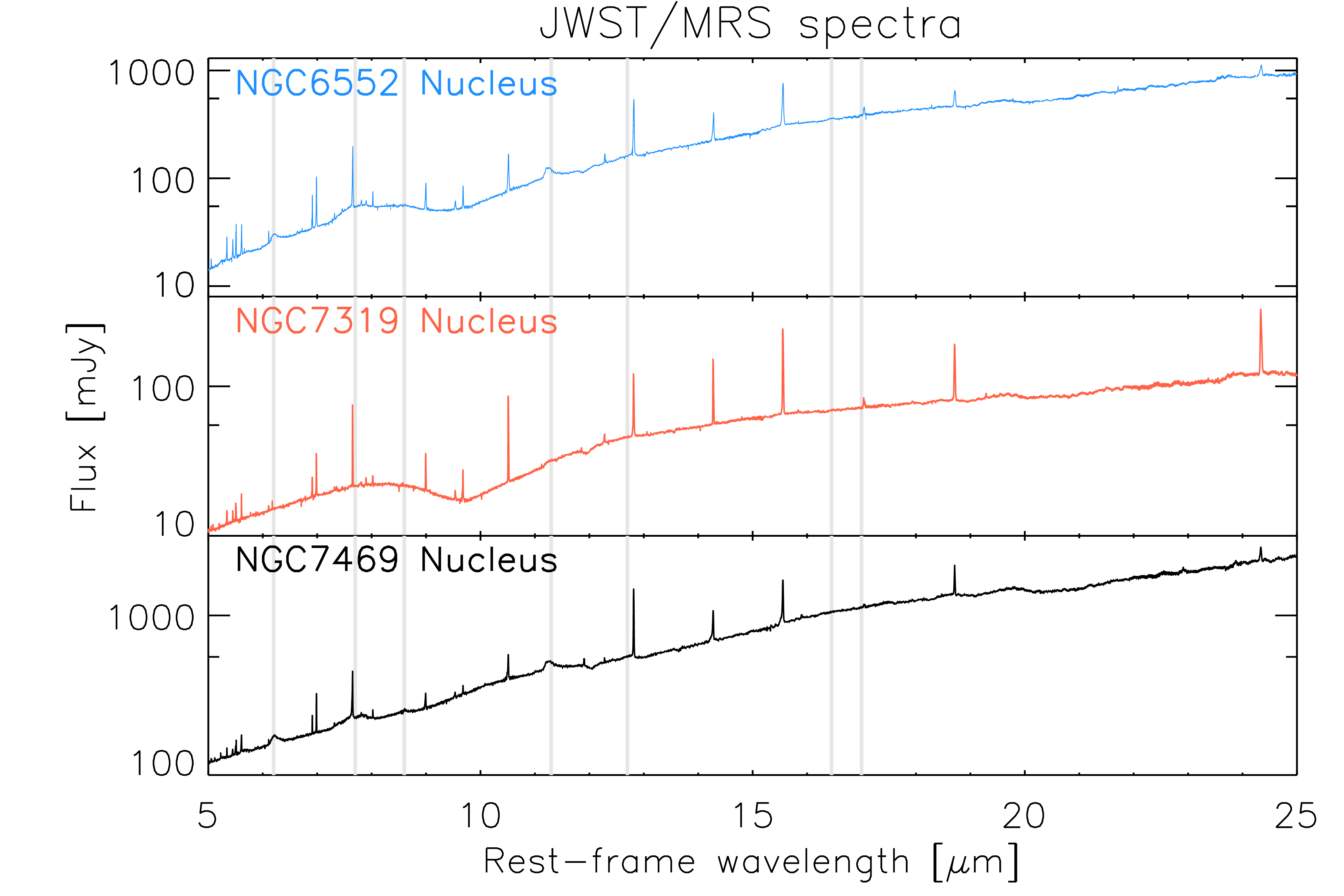}
\par}
\caption{JWST/MRS nuclear spectra of NGC\,6552, NGC\,7319, and NGC\,7469. The vertical grey lines correspond to the main PAH bands (i.e. 6.2, 7.7, 8.6, 11.3, 12.7, 16.5, and 17 $\mu$m features). }
\label{spectra}
\end{figure}

To extract the JWST/MRS spectra from the nuclear regions, we chose 1\arcsec diameter extraction apertures, which correspond to $\sim$2 times the full width half maximum of the MRS point spread function (PSF) in ch3 (see also \citealt{Pereira-Santaella22}).  For the nuclear regions of NGC\,6552, NGC\,7319, and NGC\,7469, we also applied an aperture correction using a standard star (applying the same aperture correction as described in \citealt{Pereira-Santaella22}), which is equivalent to a point source extraction. We note that the nuclear spectra of NGC\,6552, NGC\,7319, and NGC\,7469 correspond to physical scales of $\sim$245, 201, and 142 pc (at 12\,$\mu$m), respectively. The nuclear spectra of the three galaxies are shown in Fig. \ref{spectra}.

\begin{figure*}
\centering
\par{
\includegraphics[width=6.2cm]{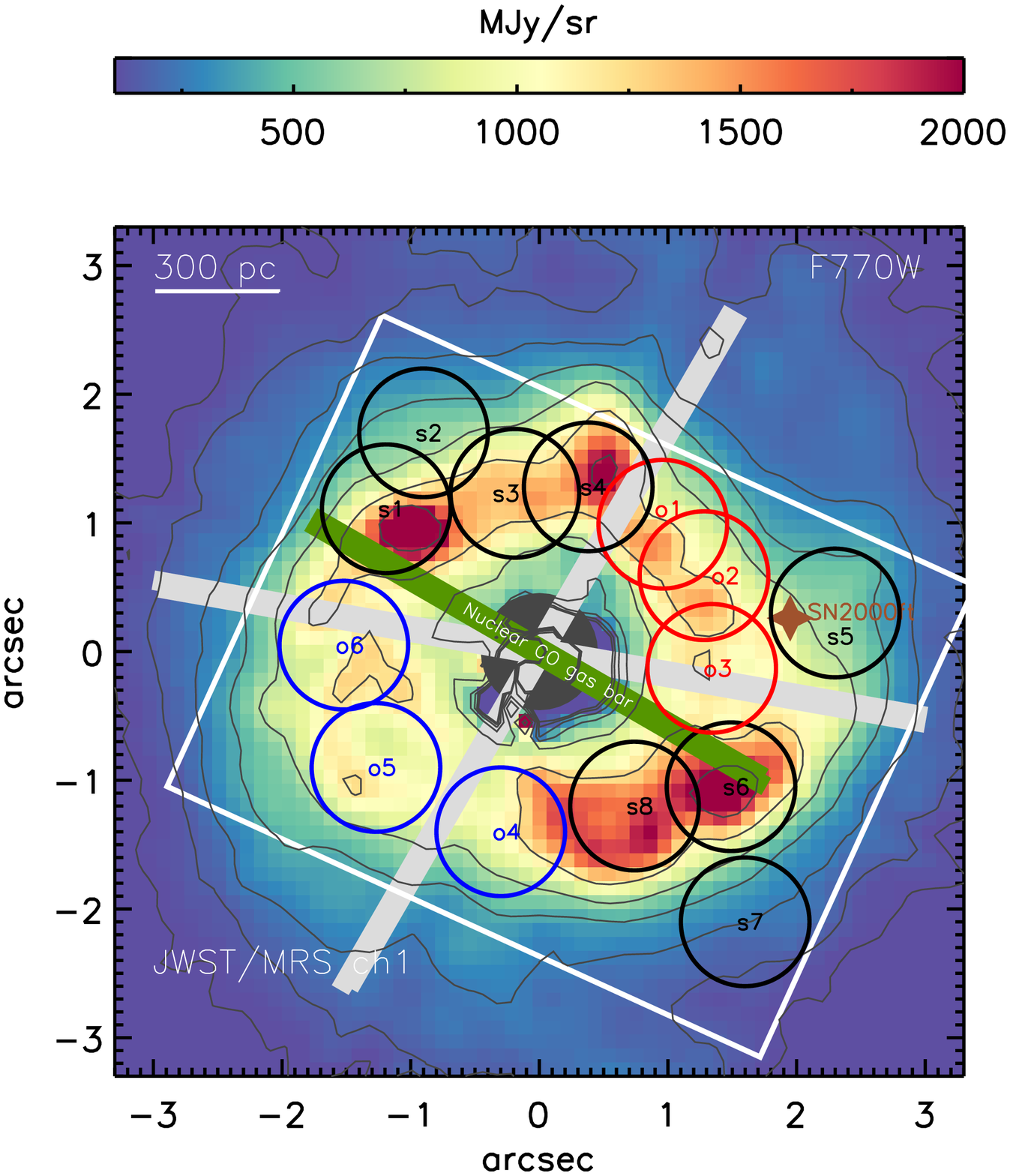}
\includegraphics[width=6.2cm]{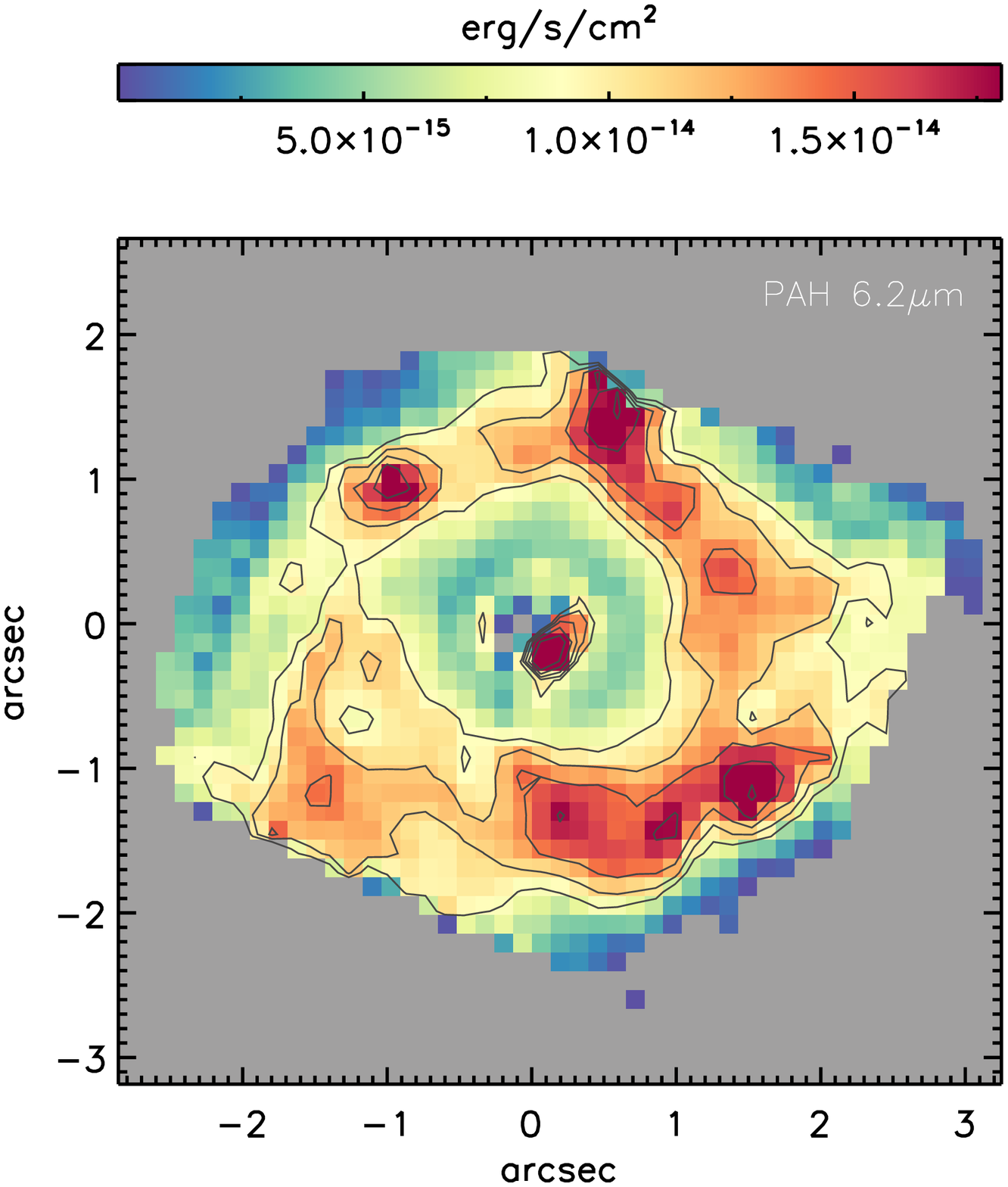}
\includegraphics[width=6.2cm]{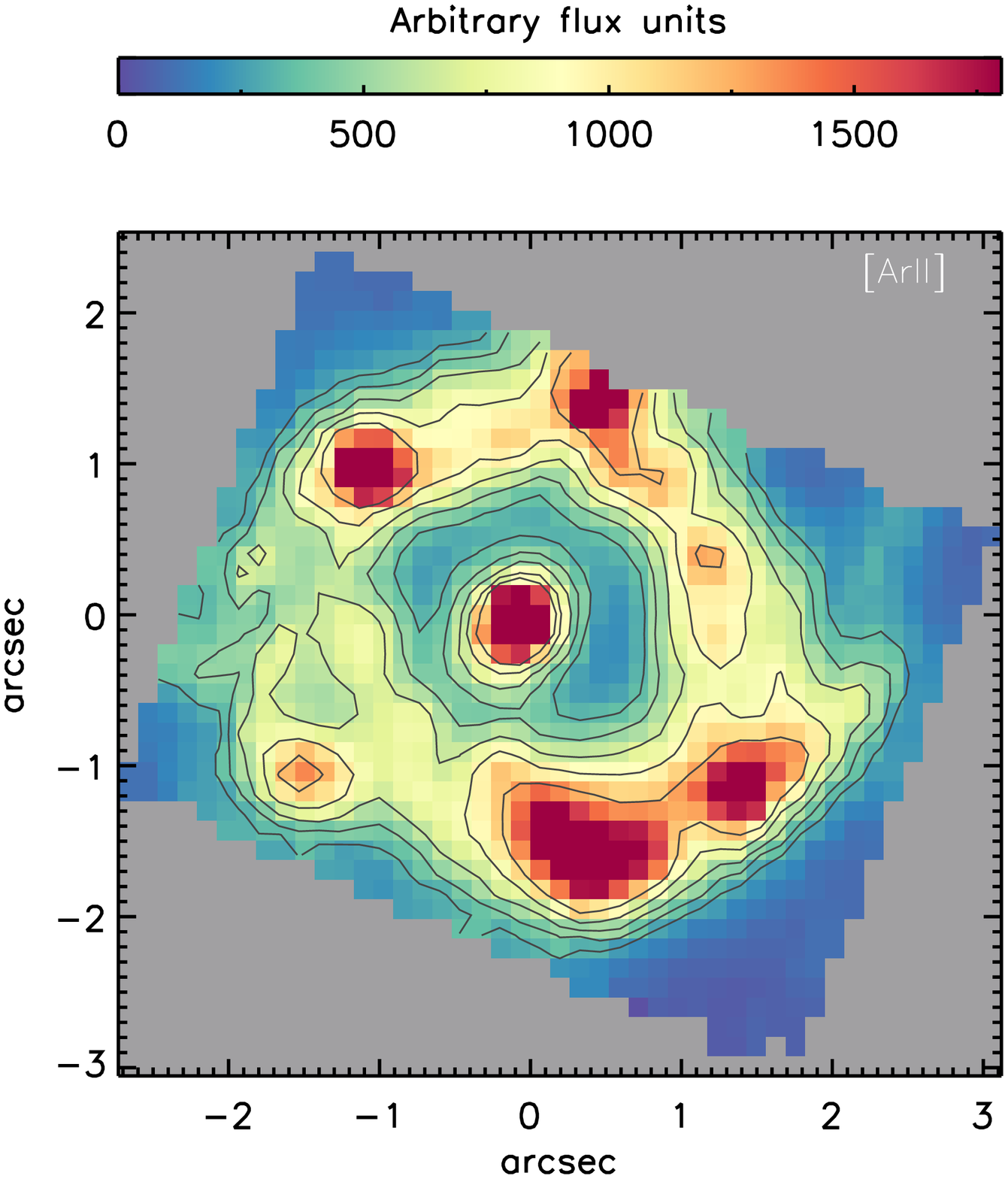}
\includegraphics[width=6.2cm]{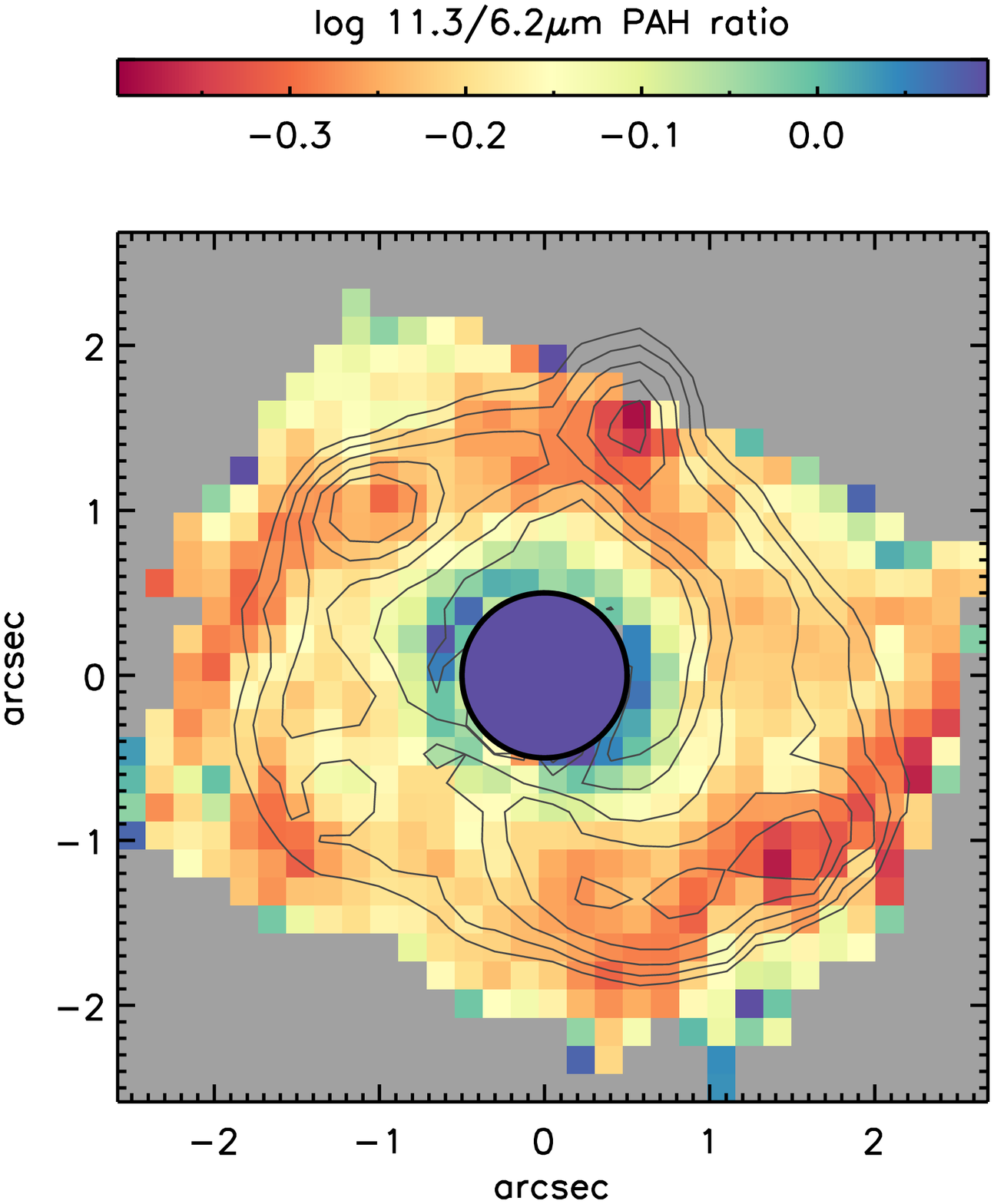}
\par}
\caption{Maps of the central $\sim$6\arcsec region of NGC\,7469, which includes the AGN and the circumnuclear ring of star formation. \textit{Top-left panel:} In colour and black contours is the JWST/F770W PSF-subtracted image (which mainly traces the 7.7 $\mu$m PAH band) . Black regions (s1, s2, s3, s4, s5, s6, and s7) correspond to selected circumnuclear zones of NGC\,7469. Red and blue regions (o1, o2, o3, o4, o5, and o6) are in the outflow region. The green line represents the orientation of the nuclear molecular gas bar. The grey lines correspond to the approximate outflow region according to the [S\,IV]$\lambda$10.51$\mu$m velocity map (see Appendix \ref{moremaps}). The white box represents the JWST/MRS ch1 FoV (3.2\arcsec $\times$ 3.7\arcsec), which is practically identical to the \textit{Spitzer}/IRS angular resolution. The brown star corresponds to the approximate location of the radio supernova SN 2000ft \citep{Colina01}. \textit{Top-right panel:} JWST/MRS 6.2 $\mu$m PAH band map derived using a local continuum (see text). \textit{Bottom-left panel:}  [Ar\,II]$\lambda$6.99$\mu$m emission map. \textit{Bottom-right panel:}  11.3/6.2 $\mu$m PAH ratio using local continua (see text). In black are the 6.2 $\mu$m PAH band contours. The central region corresponds to this PAH ratio in the nuclear spectrum. All the images are shown on a linear colour scale. North is up and east is to the left, and offsets are measured relative to the AGN.}
\label{maps1}
\end{figure*}

\subsection{Circumnuclear regions of NGC\,7469}
\label{circumn_section}
To investigate possible similarities and differences in the PAH properties between regions dominated by AGN and star formation activity, we selected a number of circumnuclear regions in NGC\,7469. We note that no circumnuclear regions were extracted for NGC\,6552 and NGC\,7319. In the case of NGC\,6552, the JWST/MRS observations of this galaxy were designed to test the MIRI detectors and no dithering was performed. Consequently, the cube is affected by hot pixels that could prove problematic for large apertures. 
In the case of NGC\,7319, PAH emission is extremely weak, as noted in  \citet{Pereira-Santaella22}.

As previously shown by \citet{Diaz-Santos07}, the SF ring of NGC\,7469 shows several bright mid-IR clumps (see also \citealt{Miles94,RamosAlmeida11} and references therein). The galaxy also has a nuclear gas bar detected in CO (2-1) with a scale similar to that of the stellar ring (e.g. \citealt{Davies04}). Indeed, \citet{Diaz-Santos07} found that two bright mid-IR clumps are located at the end of the nuclear molecular gas bar (see the top-left panel of Fig. \ref{maps1}). These authors also suggested that the location of the bright mid-IR clumps coincides with that of the inner Lindblad resonance of the larger-scale stellar bar.

Figure \ref{maps1} shows that there is  very good agreement between the brightest regions revealed by the MIRI/F770W image (which mainly traces the 7.7\,$\mu$m PAH emission), the 6.2\,$\mu$m PAH map, the clumps detected in the [Ar\,II]$\lambda$6.99$\mu$m map, and the CO(2-1) map, the last of which is reported in Fig. 2 of \citealt{Izumi20} (see also \citealt{Sanchez-Garcia22}). This indicates that the PAH emission mainly traces the SF regions, as expected.

To extract the spectra for the circumnuclear regions in NGC\,7469, we produced line and feature maps using the cubes from ch1 and ch2, which provide the highest angular resolution (Fig. \ref{maps1}). We employed a local continuum and a single Gaussian component to fit the entire cube spaxel by spaxel using a tool for fitting integral-field spectroscopy data (the {{ALUCINE tool}}; priv. communication, {\textcolor{blue}{Peralta de Arriba in prep.}}) In the case of the broad 6.2 and 11.3\,$\mu$m PAH features, we measured fluxes using the same method as in \citet{Hernan-Caballero11}. We first fitted a local continuum that we then subtracted from the observed spectra and integrated the residual data in a spectral range centred on 6.25 (from 6.0 to 6.5\,$\mu$m) and 11.3\,$\mu$m (in the range 11.05 to 11.6$\mu$m)\footnote{The 11.3\,$\mu$m PAH feature is at the end of the ch2 wavelength coverage and, thus, its local continuum may be significantly affected.}. In the case of the 7.7\,$\mu$m PAH feature, we instead used the MIRI/F770W image to study its extended morphology. This was necessary because the 7.7\,$\mu$m PAH band is split between MRS channels 1 and 2. We stress that we only used the F770W image as a proxy of the  7.7\,$\mu$m PAH feature for the morphology comparison presented in Fig. \ref{maps1}. In addition, to better trace the circumnuclear emission, we employed the MIRI/F770W PSF-subtracted image, following the the same method as described in \citet{Bernete16}.

The MIRI/MRS [S\,IV]$\lambda$10.51$\mu$m (ionisation potential, IP: 34.8\,eV) velocity and dispersion maps (shown in Appendix \ref{moremaps}) show the signatures of a potential nuclear outflow in NGC\,7469. These findings are in agreement with the work of \citet{Xu22}, who found that the H$\alpha~\lambda$6563$\AA$  and [O\,III]$\lambda$5007$\AA$ emissions are distributed asymmetrically from the north-west to the south-east region of the galaxy ($\sim$2\arcsec). They interpreted this as AGN feedback due to the outflow. Interestingly, the PAH and low ionisation potential (IP) emission line maps are relatively weak in the region that overlaps with the outflow signatures (marked in the bottom-left panel of Fig. \ref{maps1}). Indeed, the 6.2 and 7.7 $\mu$m PAH maps reveal a region about $\sim$1-1.5\arcsec (316-474\,pc) to the south-east of the active nucleus with PAH emission deficit (and low IP fine-structure lines), which is also coincident with a region of low CO(2-1) emission (see Fig. 2 of \citealt{Izumi20}). If the outflow has a wide angle, or the inclination of the outflow axis relative to the galaxy disk is small, this might be due to the clearing of PAH molecules and molecular gas by the nuclear outflow (e.g. \citealt{Herrero18,Bernete21}). However, a detailed kinematics analysis is needed to confirm this result, which is beyond the scope of this work.

Considering the morphology of the circumnuclear emission of NGC\,7469, we chose 13 regions, including emission along the SF ring, the faint emission beyond the ring, which is delimited by the edge of the MIRI field of view (FoV), and the (projected) direction of the potential outflow (see the top-left panel of Fig. \ref{maps1}).  

\begin{figure*}
\centering
\par{
\includegraphics[width=15.8cm, clip, trim=40 50 100 60]{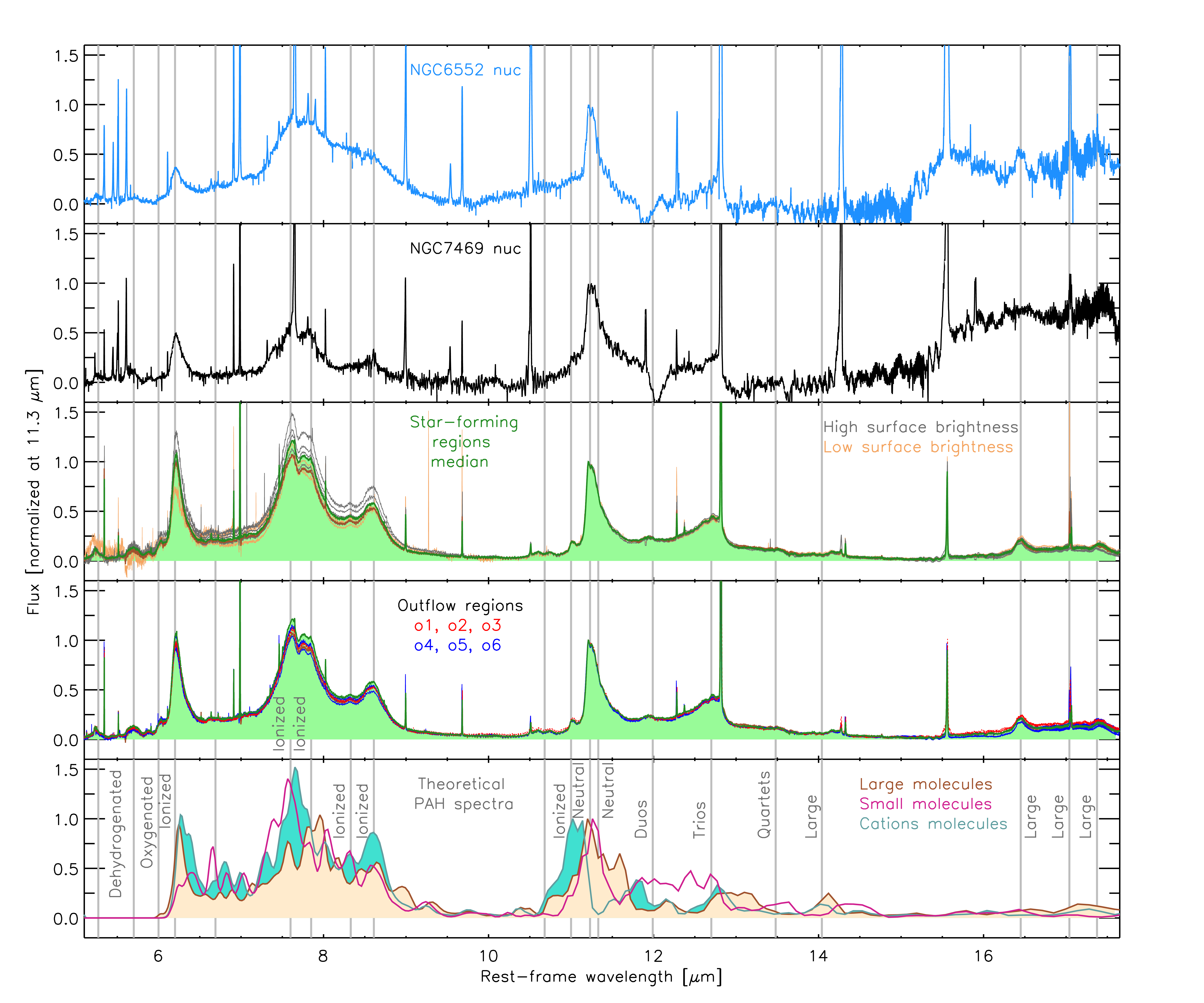}
\par}
\caption{Continuum-subtracted spectra showing the PAH emission together with the fine-structure lines. The first two panels correspond to the nuclear spectra of NGC\,6552 and NGC\,7469, respectively (see Appendix \ref{appendix_spectra} for a comparison with the PAH residual spectra of NGC\,7319). The central panel corresponds to the circumnuclear regions of NGC\,7469 (solid grey lines) and the median circumnuclear spectra of high and low surface brightness regions (solid green and brown lines, respectively). The fourth panel shows the comparison between the median NGC\,7469 circumnuclear spectrum and the regions located in the potential outflow zones (blue and red lines). The bottom panel shows the  synthetic PAH spectral templates of large (100$<$N$_{\mathrm c}<$400), small (20$<$N$_{\mathrm c}<$100), and ionised PAH molecules from \citet{Rigopoulou21}. All the spectra are normalised at 11.3 $\mu$m.}
\label{pah_profiles}
\end{figure*}

\subsection{Mid-IR modelling}

To fit the mid-IR continuum and in particular the PAH emission features, we used a modified version of \textit{PAHFIT} that includes the 10 and 18 $\mu$m silicate features in emission (e.g. \citealt{Gallimore10}) and allows for a small shift (0.1\,$\mu$m) in the fitted central PAH band wavelength ({\textcolor{blue}{Donnan in prep.}}). We masked the fine-structure lines since they are not relevant for the continuum and PAH modelling. The fits for the nuclear and circumnuclear regions are shown in Appendix \ref{appendix_spectra}.

\section{High angular and spectral resolution mid-IR spectra of AGN}

In Fig. \ref{spectra} we show the JWST/MRS nuclear spectra of NGC\,6552, NGC\,7319, and NGC\,7469, which correspond to physical scales of $\sim$245, 201, and 142\,pc, respectively. All three nuclei show the 11.3 $\mu$m PAH feature (although it is weak in the case of NGC\,7319)\footnote{Nuclear spectra show a decrease in flux at $\sim$12 $\mu$m, which is likely an instrumental artefact since the observed-frame wavelength is similar in all three galaxies and is not seen in previous ground- and space-based observations.}, supporting the notion that PAH molecules are present in the inner $\sim$142-245\,pc of the three Sy galaxies studied in this work. It is perhaps worth stressing that the detection of the 11.3 $\mu$m PAH feature is independent of the presence of a nuclear SB, as is the case for NGC\,7469.

To investigate differences in the spectra of the three galaxies, we removed the fitted continuum (see Sect. \ref{fitting}). It is evident from Fig. \ref{pah_profiles} that the spectra of NGC\,6552 and NGC\,7469 show the 6.2, 7.7, 8.6, 11.3, and 12.7\,$\mu$m PAH emission features. In the case of NGC\,7319 (see Fig. \ref{nuclear_fit2}), there is either no nuclear star formation activity or the 6.2 and 7.7\,$\mu$m PAH features have been destroyed in the inner $\sim$200\,pc (alternatively, these features could be diluted by the strong AGN continuum; e.g. \citealt{Herrero14}). 

Interestingly, all the selected SF regions in NGC\,7469 (third panel of Fig. \ref{pah_profiles}) show bright PAH emission with practically identical PAH profiles, although with some small differences in their PAH ratios (see Sect. \ref{diagram}). The strong PAH bands in these regions indicate the presence of intense star formation activity. By combining the spectra of the SF regions with high and low PAH (and [Ar\,II]) surface brightness, we produced two median circumnuclear spectra (solid green and brown lines, respectively), which we used as the SF PAH templates. The comparison of the SF PAH templates with the nuclear spectra reveals that the relative strength of the 6.2 and 7.7\,$\mu$m PAH features is lower in the nuclear region of NGC\,6552 and NGC\,7469 compared to that of the 11.3\,$\mu$m PAH feature. In the case of NGC\,7319, only the 11.3$\mu$m PAH feature is clearly detected in the nuclear region, and its continuum-subtracted residuals show tentative detection of the 12.7\,$\mu$m PAH band (see Fig. \ref{nuclear_residuals}).

On the other hand, the PAH spectra of the selected outflow regions (fourth panel of Fig. \ref{pah_profiles}) are similar to that of the low surface brightness SF template (with relatively weak 6.2 and 7.7\,$\mu$m PAH features). Indeed, the outflow regions are also coincident with regions with low surface brightness of [Ar\,II], 6.2, and 7.7 $\mu$m PAH emission (see Fig. \ref{maps1}). These regions also show low surface brightness of the CO(2-1) emission (see Fig. 2 of \citealt{Izumi20}). This indicates a higher fraction of ionised PAH molecules in the bright SF regions, which can be explained by their intense star formation activity.

\begin{figure}
\centering
\par{
\includegraphics[width=7.0cm, clip, trim=10 10 50 20]{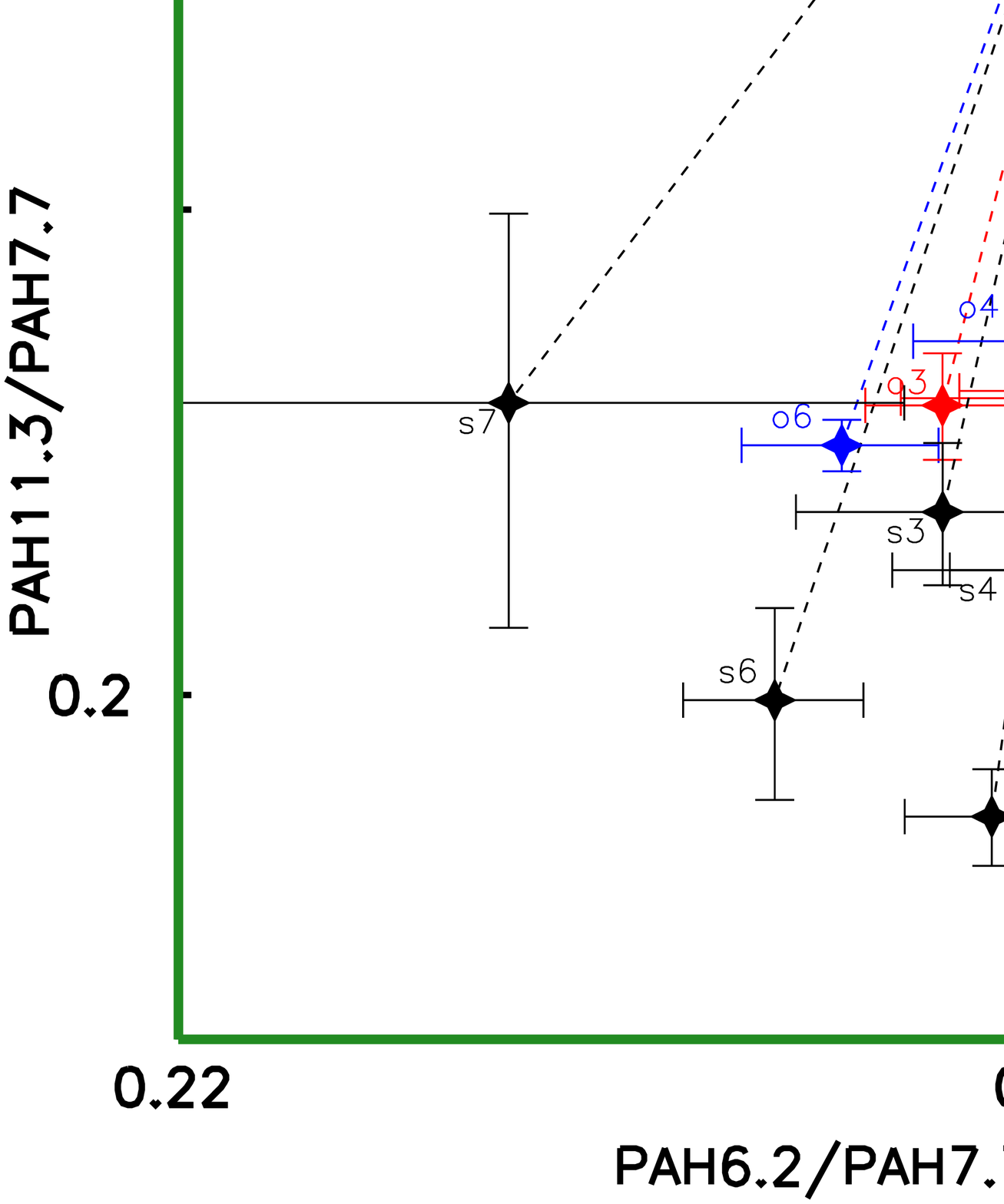}
\par}
\caption{PAH diagnostic diagram. {\it{Top panel:}} Relative strengths of the 6.2, 7.7, and 11.3~$\mu$m PAH features for NGC\,6552 (light blue circle), NGC\,7319 (solid orange line), and NGC\,7469 (black circles). Filled stars correspond to the circumnuclear emission of NGC\,7469 (coloured stars correspond to the SF regions discussed in Sect. 3.2). The purple points correspond to PAH ratios of NGC\,7469 derived from \textit{Spitzer}/IRS. The grey shaded region denotes the average location of SF galaxies from \citet{Bernete22} using \textit{Spitzer}/IRS data. The grey grid corresponds to neutral PAHs ranging from small PAHs (N$_{\mathrm C}$=20; right side of the grid) to large PAH molecules (N$_{\mathrm C}$=400; left side of the grid). Dashed grey lines correspond to intermediate numbers of carbons. The orange grid corresponds to 70\% of ionised PAH molecules for the same number of carbons as the neutral grid. Dotted orange lines correspond to intermediate numbers of carbons. The nuclear points are more consistent with neutral PAH grids. {\it{Bottom panel:}} Zoomed-in view of the green box in the top panel.}
\label{pah_diagram}
\end{figure}

The spectral resolution and sensitivity afforded by JWST/MRS has enabled the detection of even weaker PAH features in addition to the main PAH bands in the nuclear and circumnuclear regions. For instance, the 5.3 and 5.6 $\mu$m PAH features (also hinted at in previous \textit{Spitzer} observations) are clearly seen in the circumnuclear regions of NGC\,7469. These PAH bands are associated with dehydrogenated molecules according to theoretical PAH spectra (see the NASA Ames PAH IR Spectroscopic Database; \citealt{Bauschlicher18}). The 11.0 $\mu$m PAH feature, which has been associated with cation PAH molecules in galactic sources (e.g. \citealt{Shannon15}), is also clearly detected, as is the 6.0 $\mu$m PAH band, which mainly arises from oxygenated PAH molecules (e.g. \citealt{Peeters02}). Finally, previously blended PAH bands (so-called PAH complexes) in low resolution spectra are now clearly resolved, namely the 8.3, 8.6, 12.0, 16.5, and 17 $\mu$m PAH bands.

\section{Nuclear versus SF PAH properties}
\label{diagram}
Previous works established the use of the PAH band ratios as diagnostics of the physical conditions of the ISM in galaxies (e.g. \citealt{Li20} for a review). In this section we compare the relative strengths of the observed PAH bands with model grids generated using theoretically
computed PAH spectra based on density functional theory (\citealt{Rigopoulou21}). Neutral PAH molecules are responsible for the 3.3 and 11.3~$\mu$m features, whereas the 6--9~$\mu$m features originate in ionised PAH molecules (e.g. \citealt{Allamandola89,Draine07,Draine20}). In particular, the 6.2$/$7.7 and 11.3/7.7\,$\mu$m PAH ratios are sensitive to the size and ionisation stage of PAH molecules (e.g. \citealt{Rigopoulou21}).

In Fig. \ref{pah_diagram} we present, for the first time, PAH ratio diagrams involving the 6.2, 7.7, and 11.3\,$\mu$m bands and using sub-arcsecond angular resolution JWST/MRS spectra of local luminous Sy galaxies. Our first finding is that the PAH band ratios of the nuclear regions of the three Sy galaxies studied here\footnote{The 11.3/7.7\,$\mu$m PAH ratio of NGC\,7319 is an upper limit in the plot since the 6.2 and 7.7$\mu$m PAH features are not clearly detected (see Appendix \ref{appendix_spectra}).} are consistent with the presence of more neutral PAH molecules compared to those found in SF galaxies (see the top panel of Fig. \ref{pah_diagram}, where the grey area denotes the location of SF galaxies from \citealt{Bernete22}). The values of the 6.2/7.7\,$\mu$m PAH ratios for the 
nuclear region of NGC\,6552 are located in the region of large PAH molecules (N$_{\mathrm c}\sim$400; see the top panel of Fig. \ref{pah_diagram}). On the contrary, the nuclear 6.2/7.7\,$\mu$m PAH ratio of NGC\,7469 is larger than that of NGC\,6552, indicating smaller PAH molecular sizes. However, we caution that NGC\,7469 harbours a nuclear SB (\citealt{Davies04,Davies07}) and, thus, its nuclear PAH ratios could be contaminated by star formation activity.

For comparison, in Fig. \ref{pah_diagram} we also show the PAH ratios corresponding to the central and extended regions of NGC\,7469 derived from \textit{Spitzer}/IRS (purple points). At the angular resolution of \textit{Spitzer} ($\sim$4\arcsec) it is not possible to distinguish between nuclear and extended SF emission. However, JWST/MRS enables us to isolate the nuclear region of NGC\,7469, which shows a higher 11.3/7.7\,$\mu$m PAH ratio, suggesting the presence of a larger fraction of neutral molecules compared to the SF regions. This is consistent with the trend in the 11.3/6.2\,$\mu$m PAH ratio map (also a good PAH ionisation fraction indicator) shown in bottom-right panel of Fig. \ref{maps1}. This is broadly in agreement with IRS results for AGN-dominated sources (\citealt{Diamond10,Bernete22}).

A number of possibilities could account for the larger fraction of neutral PAH molecules in the nuclear region of AGN: (1) neutral PAH molecules could be protected (shielded) by the high hydrogen column density (e.g. \citealt{Herrero20} and references therein); (2) the AGN is not significantly contributing to the photoionisation of the PAH emission (e.g. \citealt{Bilton16}); and (3) ionised PAH molecules could be destroyed more easily than neutral ones (e.g. \citealt{Chabot19,Li22}). 

Based on the various SF regions of NGC\,7469, we do not find evidence for shielding of the PAH molecules by the molecular gas. This can be related to the relatively large apertures employed in this work where the hydrogen column density is likely to be very similar for the various selected SF regions. On the other hand, the hydrogen column density in the nuclear region of NGC\,7469 (N$_{\mathrm H2}=2.3\times10^{23}$\, cm$^{-2}$ estimated using the CO(1-0) at 115.27\,GHz flux from \citealt{Izumi20} and a Galactic conversion factor) is higher than those of the SF regions distributed along the ring (e.g. \citealt{Izumi20}). This suggests that shielding could, in fact, be playing a role in the nuclear region of the AGN, while this is not clear in SF regions. We also find that the regions located in the (projected) direction of the outflow (blue and red symbols in the bottom panel of Fig. \ref{pah_diagram}) have slightly larger fractions of neutral molecules (higher 11.3/7.7 $\mu$m PAH ratios) than the majority of the SF regions.

This initial analysis shows the potential of JWST/MRS to study the properties of PAH molecules on sub-arcsecond scales and investigate variations across entire galaxies.  These results demonstrate that the unique combination of high sensitivity and high spatial and spectral resolution allows the characterisation of the PAH properties in the nuclear and circumnuclear regions in local luminous Sy galaxies.

\section{Summary and conclusions}
\label{conclusions}
The advent of JWST has ushered in a new era in studies of the central regions of local galaxies. In this work we have analysed publicly available JWST MIRI/MRS data for three Sy galaxies (NGC\,6552, NGC\,7469, and NGC\,7319) and investigated the properties of their PAH emission at sub-arcsecond scales ($\sim$0.45\arcsec; physical scales $\sim$142-245\,pc). For each of the galaxies, we studied PAH spectra of their nuclear regions. In the case of NGC\,7469 we also studied the PAH emission originating in the circumnuclear SF ring. All the selected SF regions in NGC\,7469 show bright PAH emission with practically identical PAH profiles. We also find that previously blended PAH bands (8.3, 8.6, 12.0, 16.5, and 17 $\mu$m PAH features) in low resolution spectra are now clearly resolved in the SF ring. 

Our analysis allowed, for the first time, to derived the properties of the PAH molecules in the inner $\sim$ 142-245\,pc of luminous Sy galaxies (L$_{\rm bol}>10^{44.46}$\,erg/s). We find that the hard radiation fields present in the vicinity of an AGN likely have a significant impact on the ionisation state (and size) of the PAH grains.

We find that in all three cases, the nuclear PAH emission mainly originates in neutral PAHs. In contrast, PAH emission originating in the SF region favours smaller, ionised PAH grains. A number of reasons could be responsible for the differences in PAH properties between nuclear and circumnuclear regions: shielding of neutral PAHs by the high hydrogen column densities, destruction of ionised PAH grains by the central AGN, or a lack of significant AGN contribution to the photoionisation of PAHs in the innermost regions of these galaxies.

In the near future, detailed studies such as this one -- that take advantage of the unprecedented capabilities of JWST and are carried out not only with larger samples of galaxies but covering wider luminosity, hydrogen column density, and Eddington ratio ranges -- are needed to improve the statistic of the results reported in this work.

\begin{acknowledgements}
IGB and DR acknowledge support from STFC through grant ST/S000488/1. DR acknowledges support from the University of Oxford John Fell Fund. A.L acknowledges support by grant PIB2021-127718NB-100 by the Spanish Ministry of Science and Innovation/State Agency of Research (MCIN/AEI). AAH, LPdA and SGB acknowledge support from  grant PGC2018-094671-B-I00 funded by MCIN/AEI/ 10.13039/501100011033 and by ERDF A way of making Europe. S.G.B. acknowledges support from the research project PID2019-106027GA-C44 of the Spanish Ministerio de Ciencia e Innovaci\'on. EB acknowledges the Mar\'{\i}a Zambrano program of the Spanish Ministerio de Universidades funded by the Next Generation European Union and is also partly supported by grant RTI2018-096188-B-I00 funded by MCIN/AEI/10.13039/501100011033. CRA acknowledges support from grant PID2019-106027GB-C42, funded by MICINN-AEI/10.13039/501100011033, from EUR2020-112266, funded by MICINN-AEI/10.13039/501100011033 and the European Union NextGenerationEU/PRTR, and from the Consejer\'ia de Econom\'ia, Conocimiento y Empleo del Gobierno de Canarias and the European Regional Development Fund (ERDF) under grant ProID2020010105, ACCISI/FEDER, UE. The authors also thank J. {\'A}lvarez-M{\'a}rquez for discussions.

This work is based on observations made with the JWST. The authors acknowledge the MIRI comissioning, ERO and ERS teams for developing their observing program with a zero--exclusive--access period. The authors are extremely grateful to the JWST helpdesk for their constant and enthusiastic support. Finally, we thank the anonymous referee for their useful comments.
\end{acknowledgements}


\begin{appendix}

\section{Data reduction}
\label{reduction}

We retrieved mid-IR (4.9-28.1 $\mu$m) integral-field spectroscopy data observed using MIRI (\citealt{Rieke15})/MRS (\citealt{Wells15}) on the 6.5m {JWST}. The MRS comprises four wavelength channels: ch1 (4.9--7.65~$\mu$m), ch2 (7.51--11.71~$\mu$m), ch3 (11.55--18.02~$\mu$m), and ch4 (17.71--28.1~$\mu$m). These channels are further subdivided into three sub-bands (short, medium, and long). The FoV is larger for longer wavelengths: ch1 (3.2\arcsec $\times$ 3.7\arcsec), ch2 (4.0\arcsec $\times$ 4.7\arcsec), ch3 (5.2\arcsec $\times$ 6.1\arcsec), and ch4 (6.6\arcsec $\times$ 7.6\arcsec). We refer the reader to \citet{Rigby20} for further details on the JWST.

We downloaded the JWST/MRS level-2b calibrated data of NGC\,6552, NGC\,7319, and NGC\,7469 from the JWST archive. We note that the NGC\,6552 data were reduced using context 0913 of the Calibration References Data System (CRDS), whereas for NGC\,7319 and NGC\,7469 context 0925 of the CRDS was employed. Before building the 3D sub-channel cubes, we applied a residual fringe correction in the detector plane\footnote{{\textcolor{blue}{https://jwst-pipeline.readthedocs.io/en/latest/jwst/residual$\_$fringe/main.html}}} using the JWST calibration pipeline (release 1.6.2). This step corrects the low frequency fringe residuals remaining after the standard pipeline fringe flat correction. We refer the reader to \citet{Argyriou20} for details of the residual fringe correction. Following that step, we built the twelve 3D spectral sub-cubes (four MRS channels with short, medium, and long bands) for the background and science observations. Finally, we estimated the background emission by calculating the median value for each wavelength channel in their background dedicated data cubes and manually subtracted these background values from the science data cubes. For NGC\,6552, the MRS observations were designed to test the MIRI detectors, and the setups used for the background observations are not the same as those of the galaxy (see \citealt{Marquez22} for further details on these observations). By comparing the expected background\footnote{\textcolor{blue}{https://jwst-docs.stsci.edu/jwst-general-support/jwst-background-model}} of the three galaxies, we find that the backgrounds of NGC\,6552 and NGC\,7319 are very similar. Thus, we used the dedicated background of NGC\,7319 for the data cubes of NGC\,6552. 

We also compiled the fully reduced MIRI/F770W imaging data of NGC\,7469 (Program ID 1328, PI L. Armus) and the PSF standard star (BD$+$60 1753; Program ID 1027, PI M. Garc\'ia Mar\'in) using context 0913 of the CRDS. The MIRI imager sub-array SUB256 (FoV: 28.2\arcsec~$\times$~28.2\arcsec) was used for these observations.

\section{Ionised gas maps of NGC\,7469}
\label{moremaps}
In Fig. \ref{maps2} we show a comparison between the velocity and velocity dispersion maps of [S\,IV]$\lambda$10.51$\mu$m and H$_2$~(S4) at 8.03$\mu$m. [S\,IV] is a good tracer of the AGN, whereas the H$_2$~(S4) rotational line traces the warm molecular gas. Figure \ref{maps2} shows an excess in the velocity and velocity dispersion maps of the [S\,IV] emission line in the south-east region compared to that of the warm molecular gas that can be related with the nuclear outflow of NGC\,7469. Since the current wavelength calibration needs to be refined, to produce the velocity maps we first set the zero of the H$_2$~(S4) velocity map at the nuclear position. Then, for the [S\,IV] velocity field we adjusted the velocity to match the rotation of H$_2$~(S4) in regions with no clear evidence of outflow.

\begin{figure}
\centering
\par{
\includegraphics[width=8.7cm]{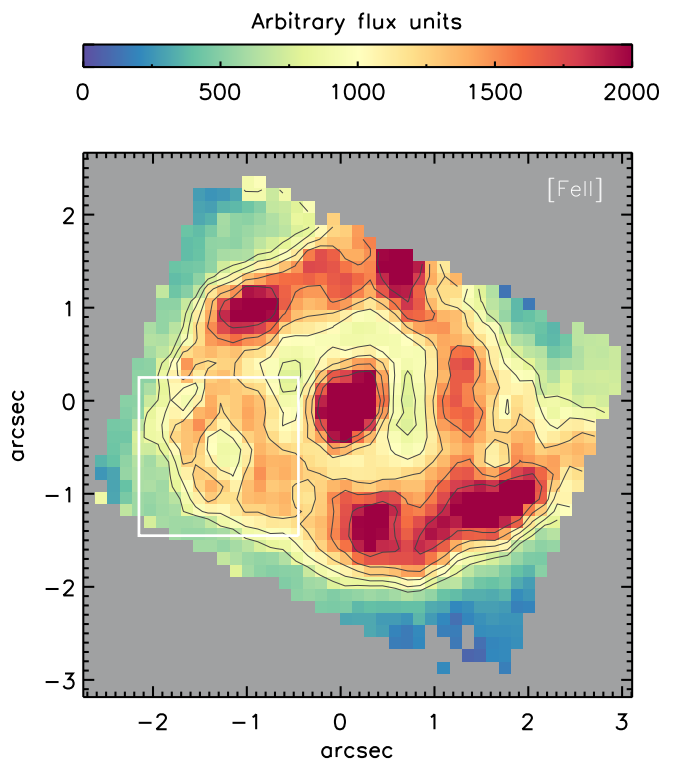}
\par}
\caption{[Fe\,II]$\lambda$5.34$\mu$m intensity map (in colour and black contours). The white box represents the region of the arc-like morphology. The map is shown on a linear colour scale. North is up and east is to the left, and offsets are measured relative to the AGN.}
\label{maps3}
\end{figure}

\begin{figure*}[h!]
\centering
\par{
\includegraphics[width=6.0cm]{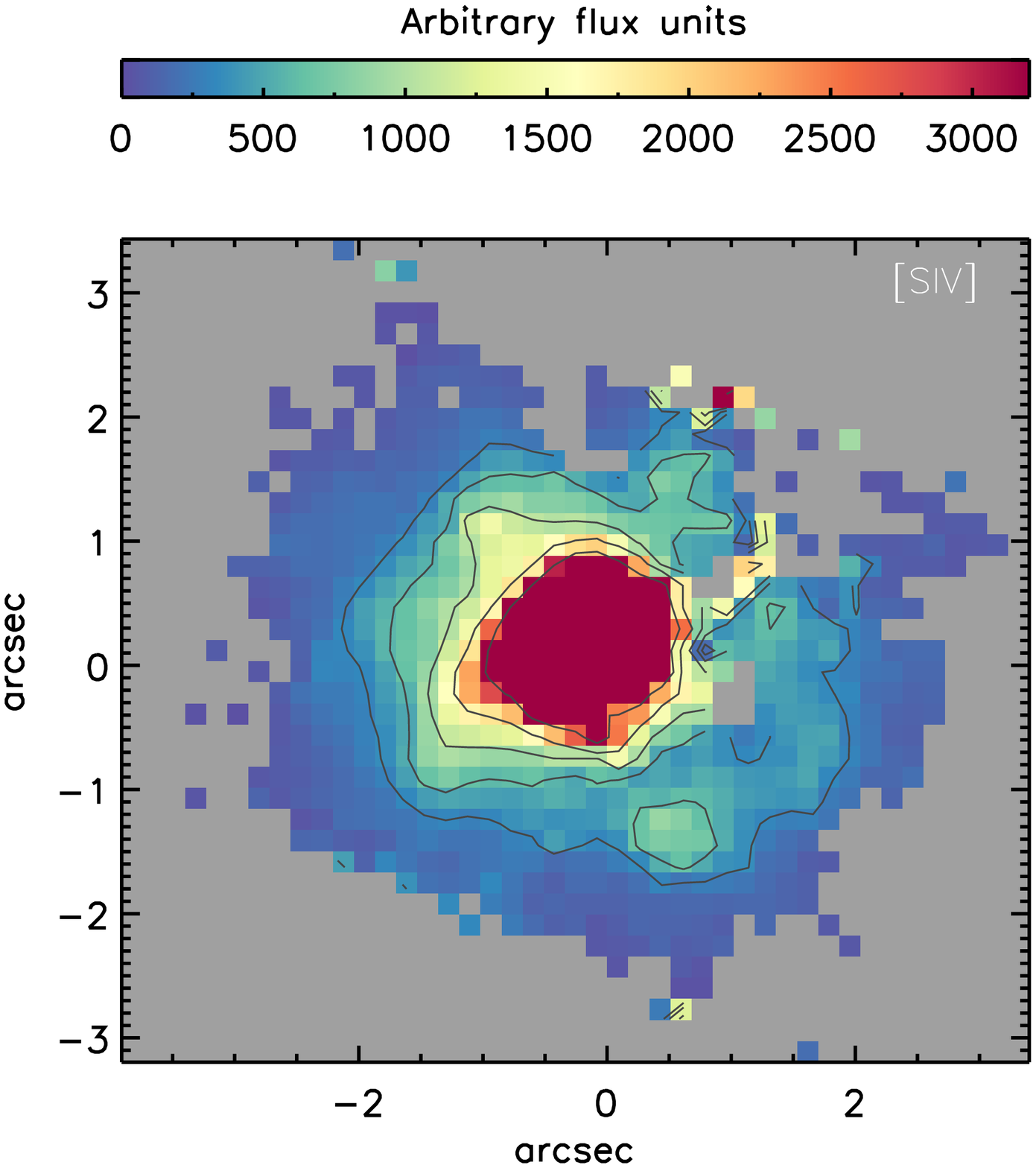}
\includegraphics[width=6.0cm]{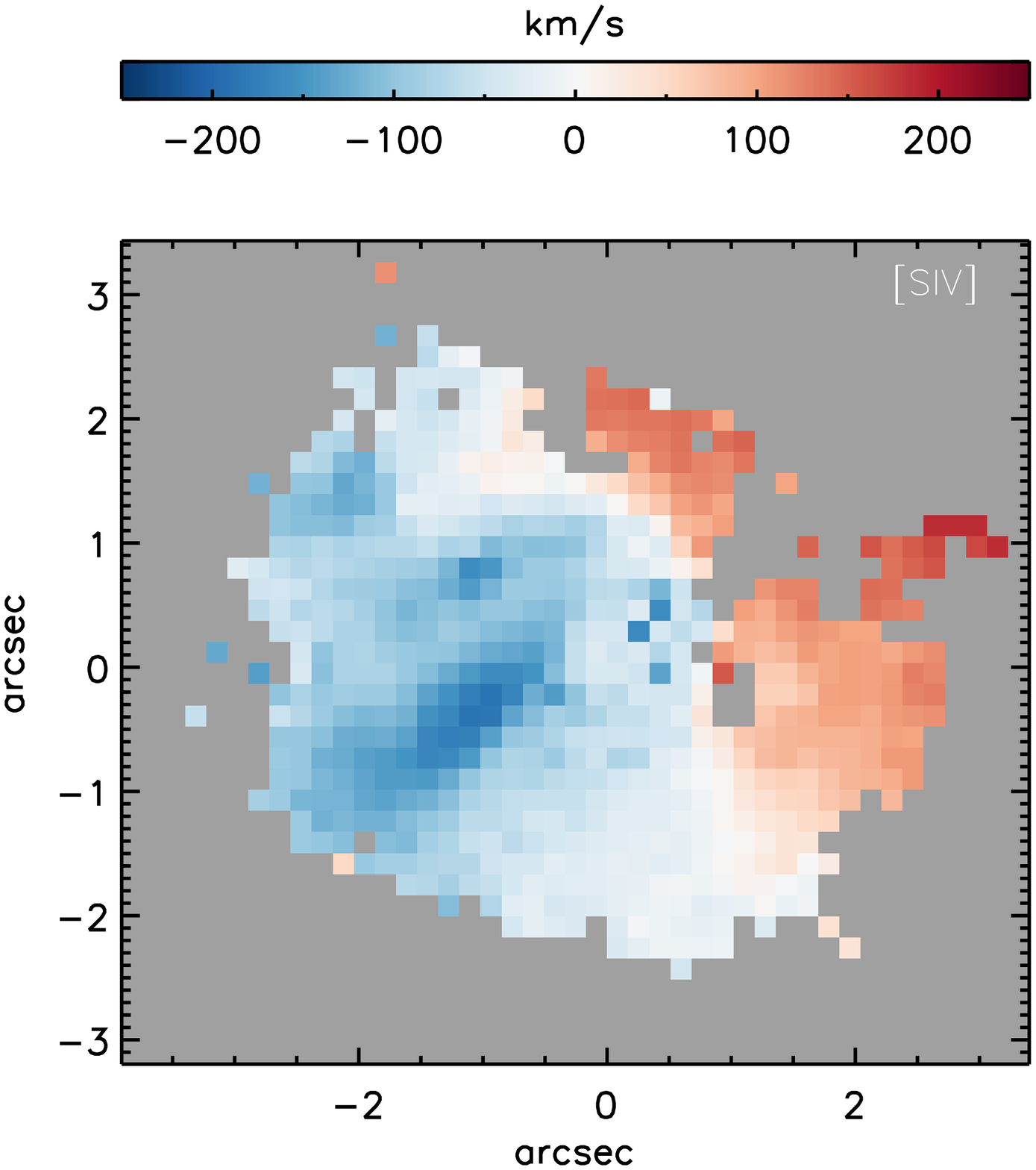}
\includegraphics[width=6.0cm]{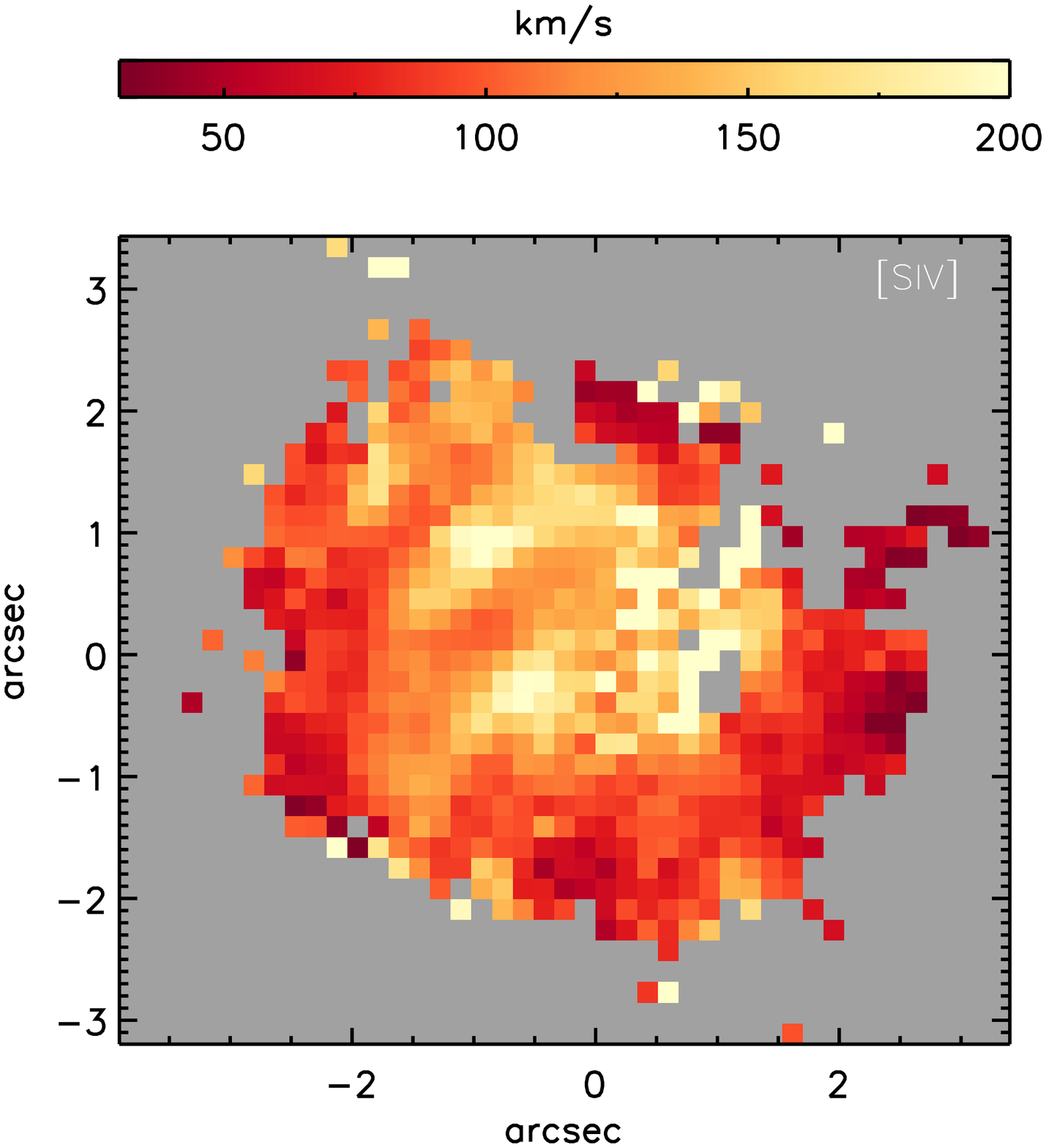}
\includegraphics[width=6.0cm]{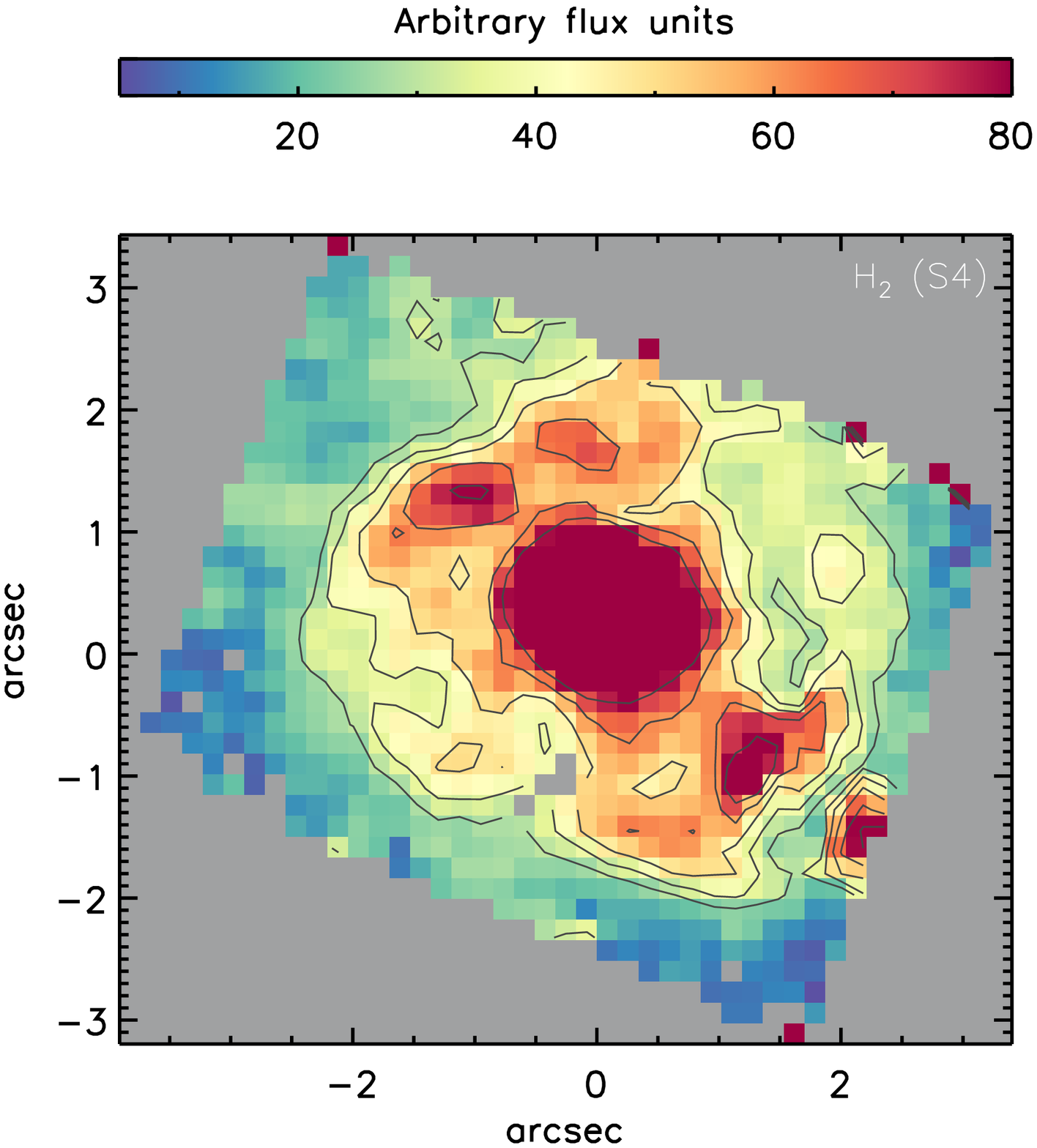}
\includegraphics[width=6.0cm]{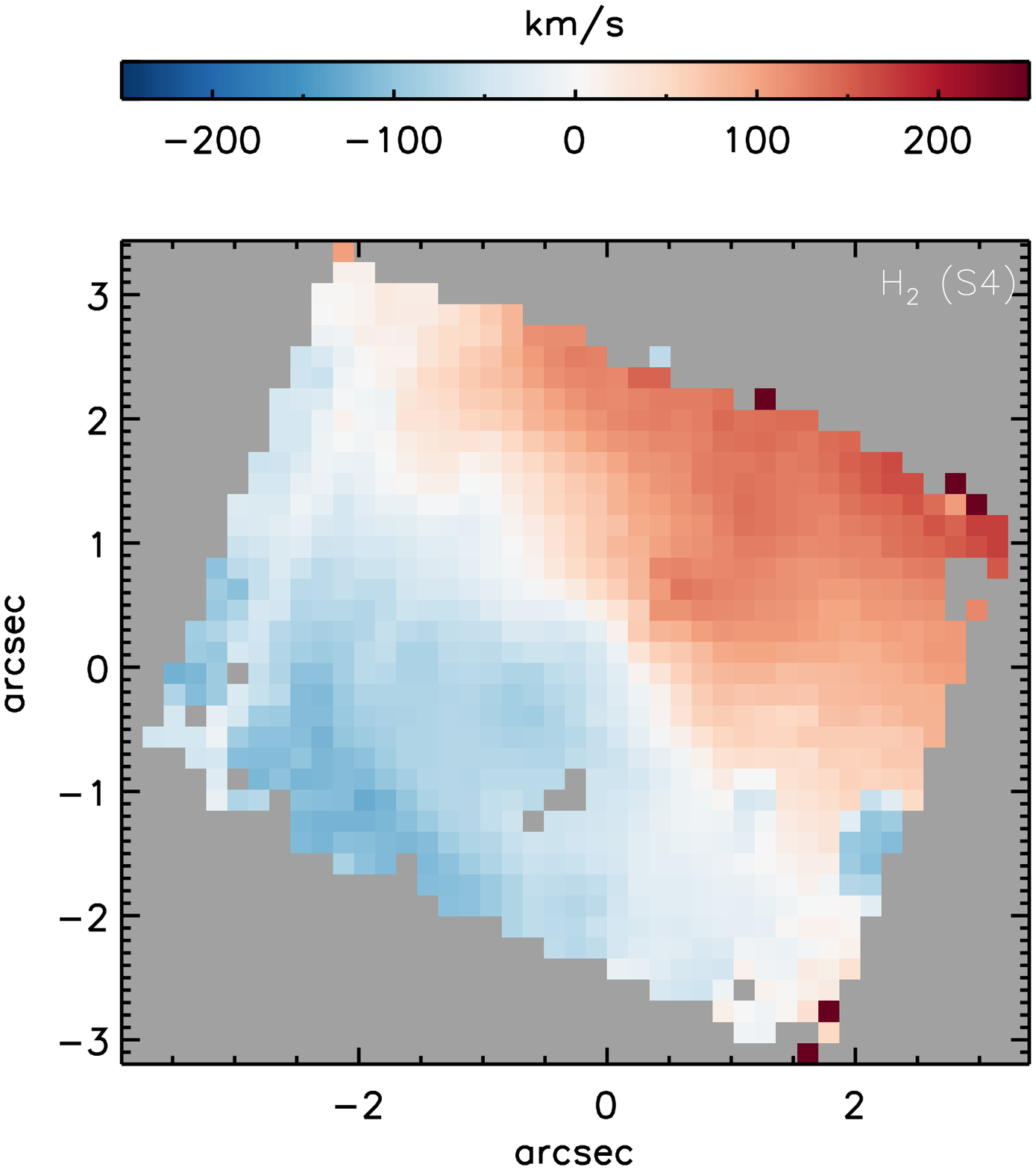}
\includegraphics[width=6.0cm]{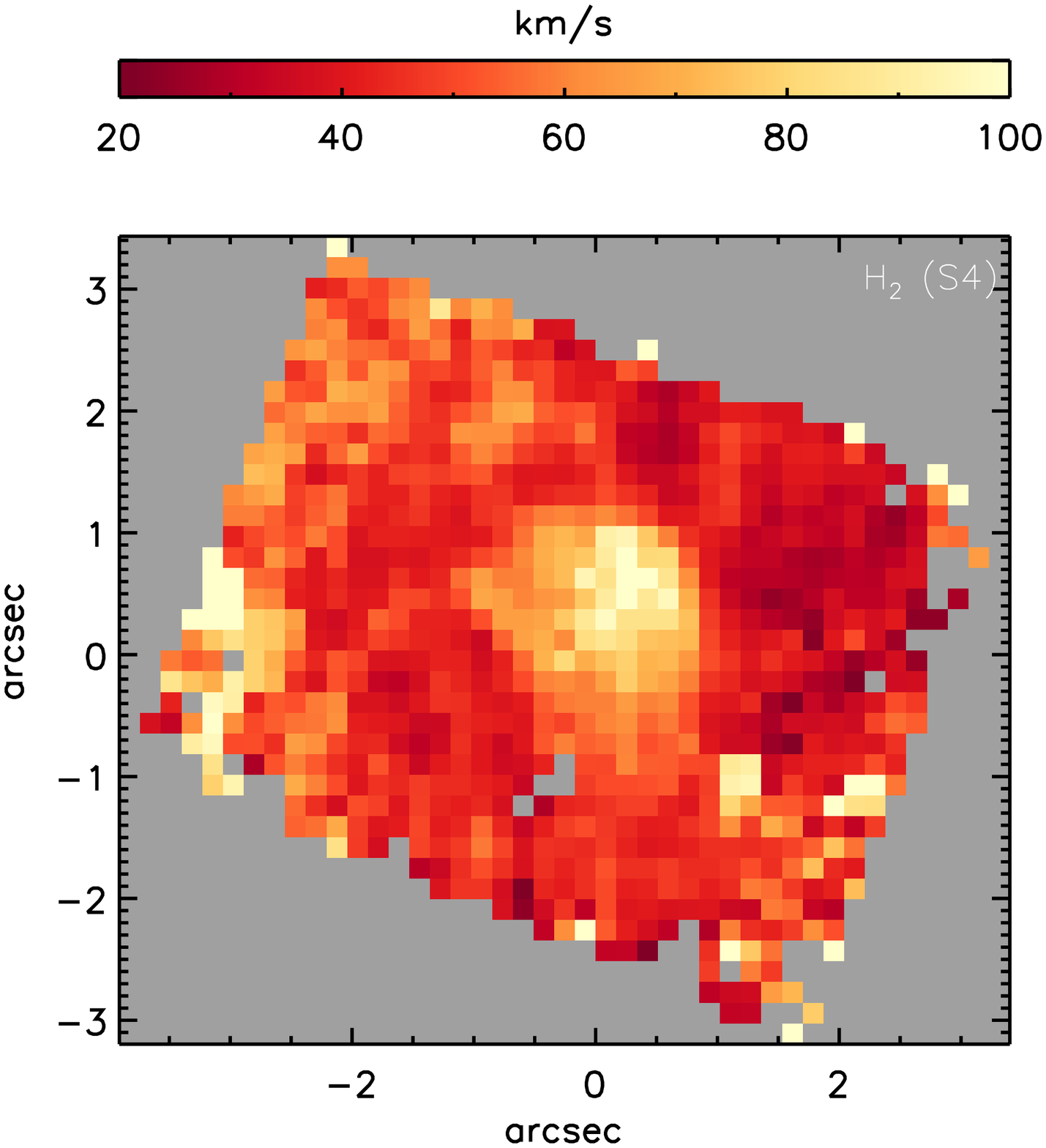}
\par}
\caption{Emission line maps. {\textit{Top-left panel:}} JWST/MRS [S\,IV]$\lambda$10.51$\mu$m intensity map. {\textit{Top-central panel:}} JWST/MRS [S\,IV]$\lambda$10.51$\mu$m velocity map. {\textit{Top-right panel:}} JWST/MRS [S\,IV]$\lambda$10.51$\mu$m velocity dispersion map. {\textit{Bottom-left panel:}} JWST/MRS H$_2$ (S4) at 8.03$\mu$m intensity map. {\textit{Bottom-central panel:}} JWST/MRS H$_2$ (S4) velocity map. {\textit{Bottom-right panel:}} JWST/MRS H$_2$ (S4) velocity dispersion map. All the maps are shown on a linear colour scale. North is up and east is to the left, and offsets are measured relative to the AGN.}
\label{maps2}
\end{figure*}

The [Fe\,II]$\lambda$5.34$\mu$m intensity map (Fig. \ref{maps3}), which is a good shock tracer, suggests there is an arc-like (or broken-ring) morphology surrounding the region potentially cleared by the nuclear outflow. The [Ar\,II]$\lambda$6.99$\mu$m map shows a concentrated clumpy emission in the southern part of the arc-like region (see Fig. \ref{maps1}). Although a detailed kinematics analysis is needed to confirm that this is indeed the case, this scenario is consistent with positive AGN feedback, which has been identified in other Sy galaxies (e.g. \citealt{Cresci15}). 

\section{Nuclear and circumnuclear mid-IR spectra modelling}
\label{appendix_spectra}
\begin{figure}[h!]
\centering
\par{
\includegraphics[width=9.8cm]{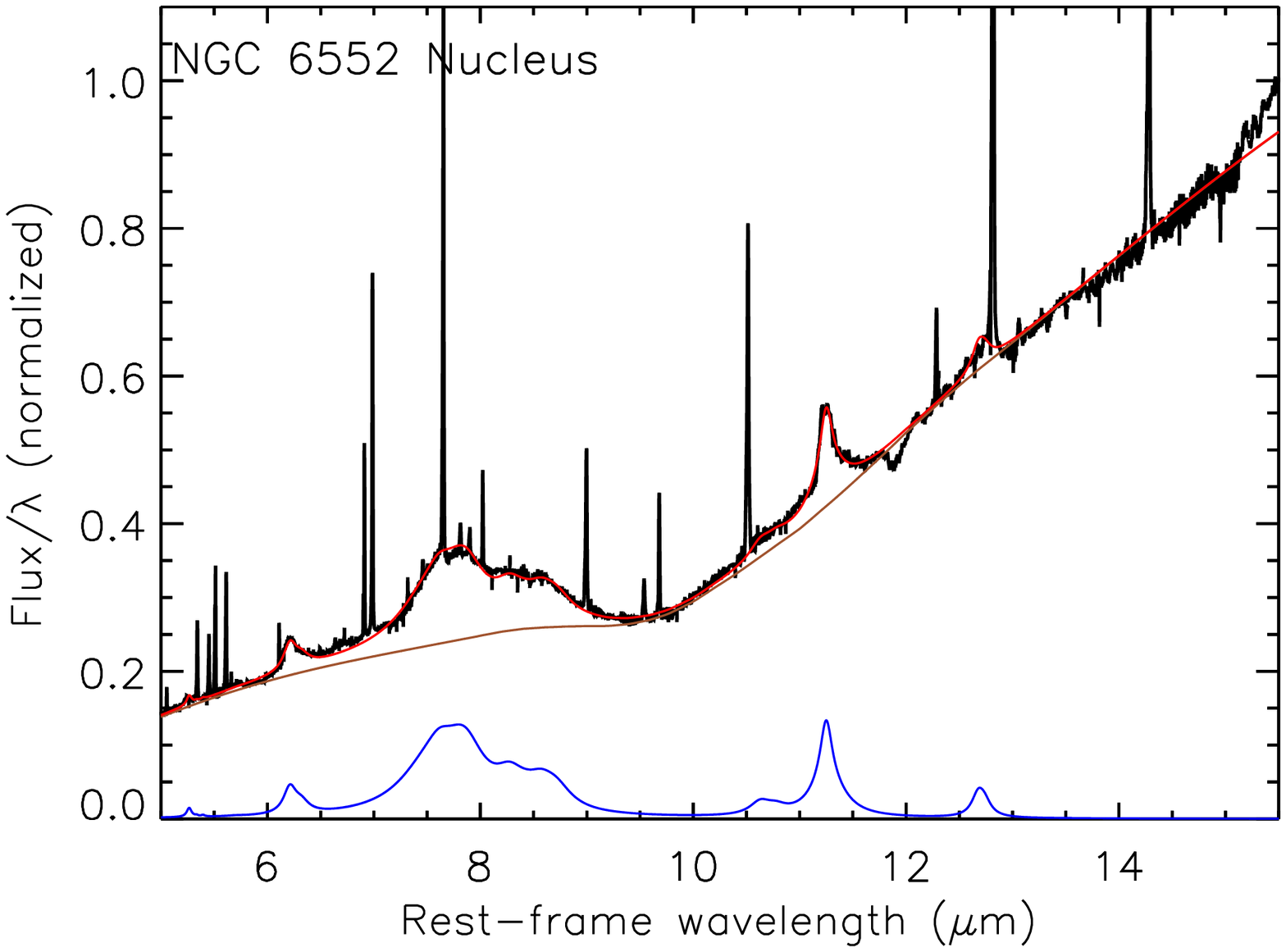}
\includegraphics[width=9.8cm]{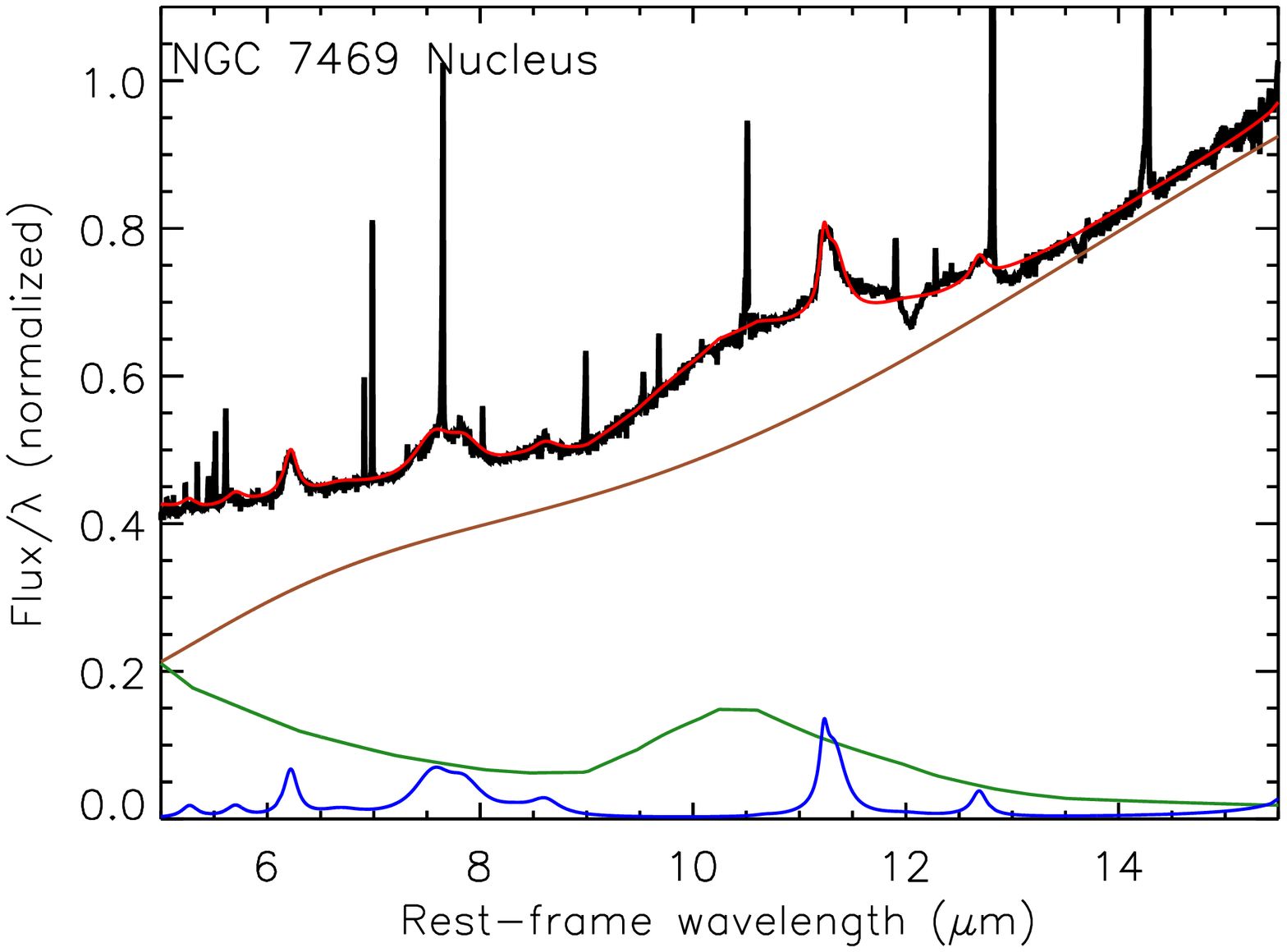}
\par}
\caption{Mid-IR spectral modelling of the nuclear regions of NGC\,6552 (top panel) and NGC\,7469 (bottom panel) using PAHFIT. The JWST/MRS rest-frame spectra and model fits correspond to the solid black and red lines. We show the continuum (solid brown lines) and the fitted PAH features (solid blue lines). The silicate feature in emission (solid green line) is included for the fit of NGC\,7469.}
\label{nuclear_fit}
\end{figure}

\begin{figure}
\centering
\par{
\includegraphics[width=9.8cm]{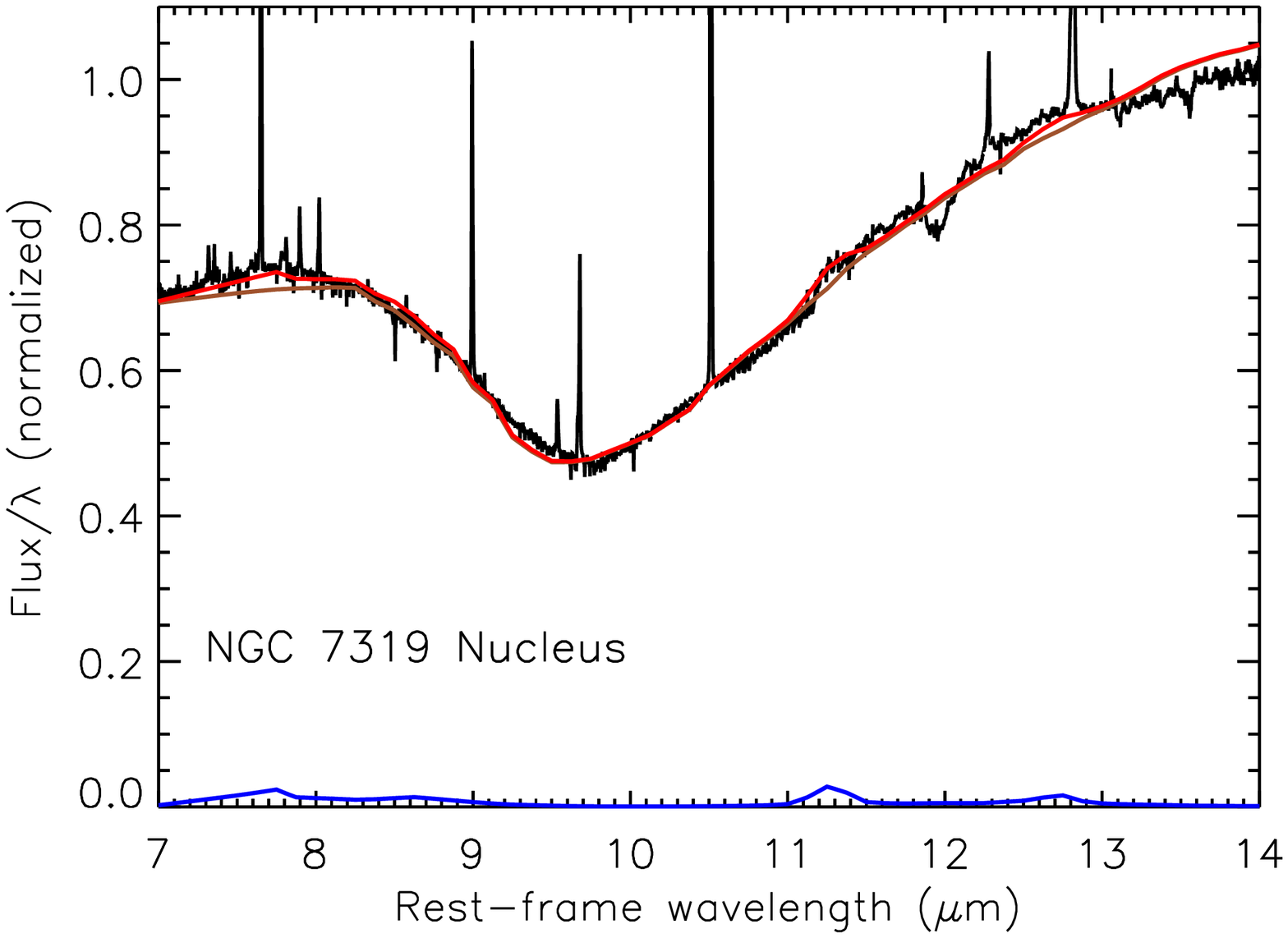}
\par}
\caption{Mid-IR spectral modelling of the nuclear regions of NGC\,7319 following the same method as in \citet{Roche07} (see text). The JWST/MRS rest-frame spectra and model fits correspond to the solid black and red lines. We show the continuum (solid brown lines) and the fitted PAH features (solid blue lines).}
\label{nuclear_fit2}
\end{figure}

\begin{figure*}
\centering
\par{
\includegraphics[width=4.8cm]{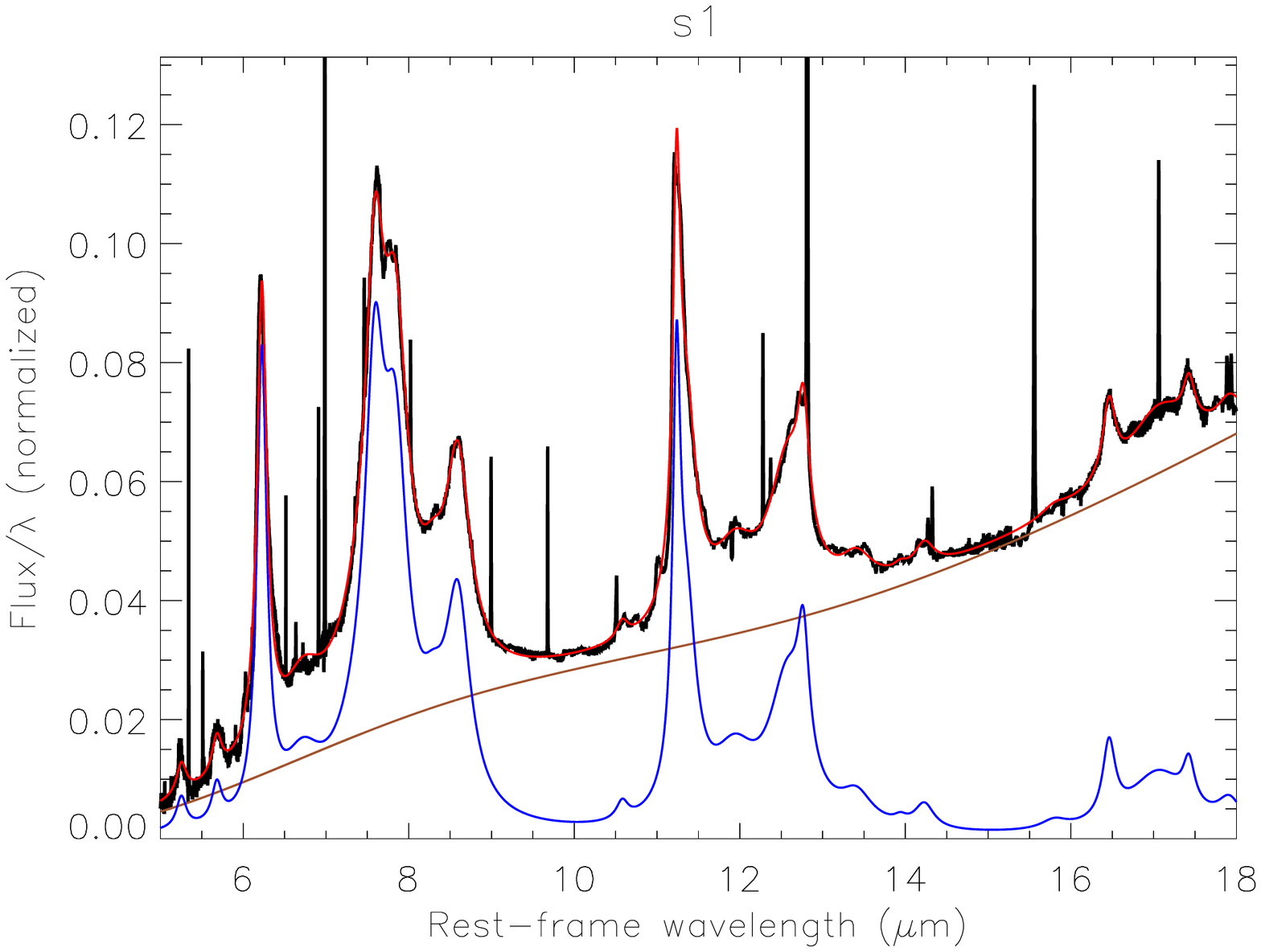}
\includegraphics[width=4.8cm]{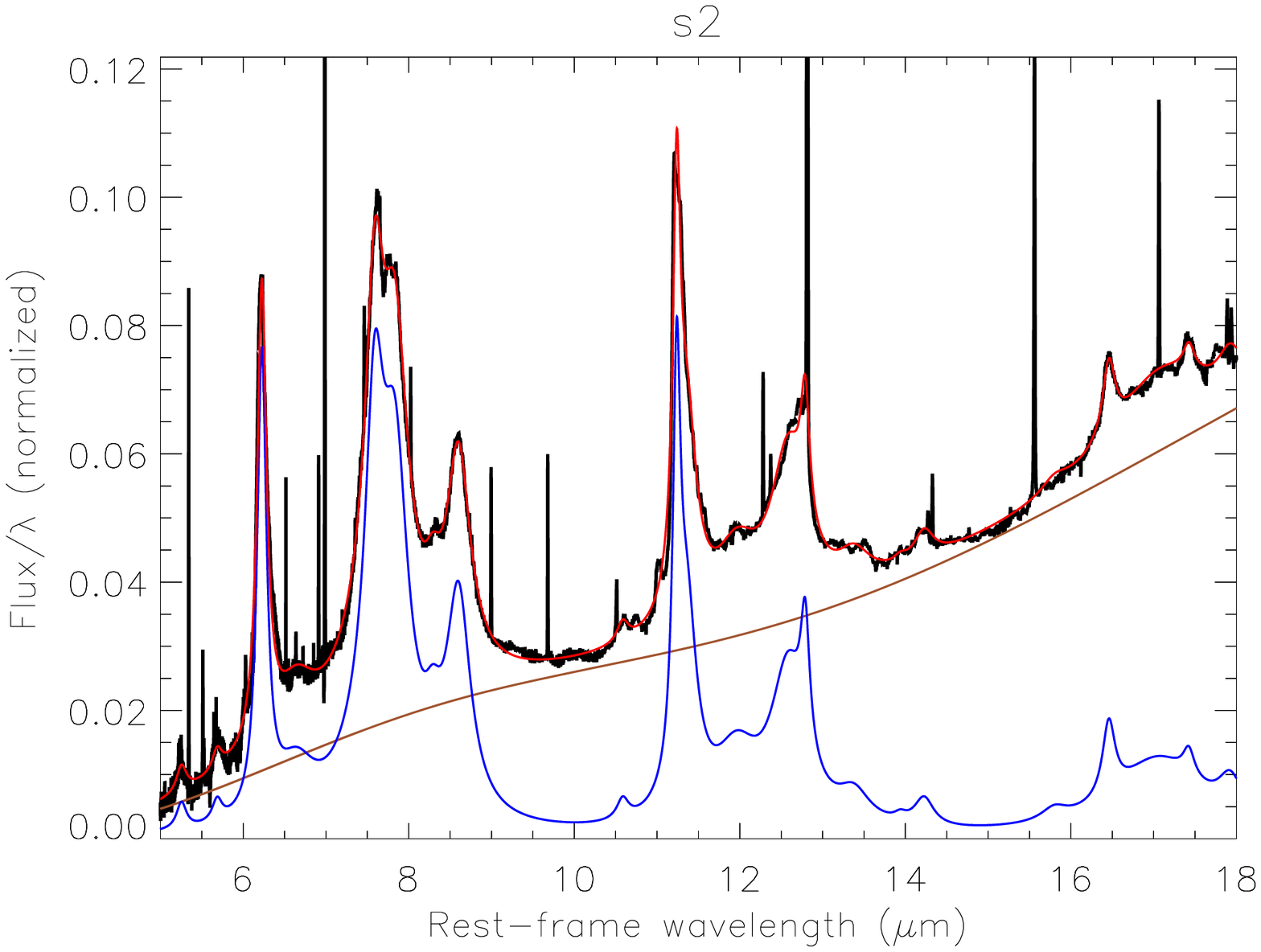}
\includegraphics[width=4.8cm]{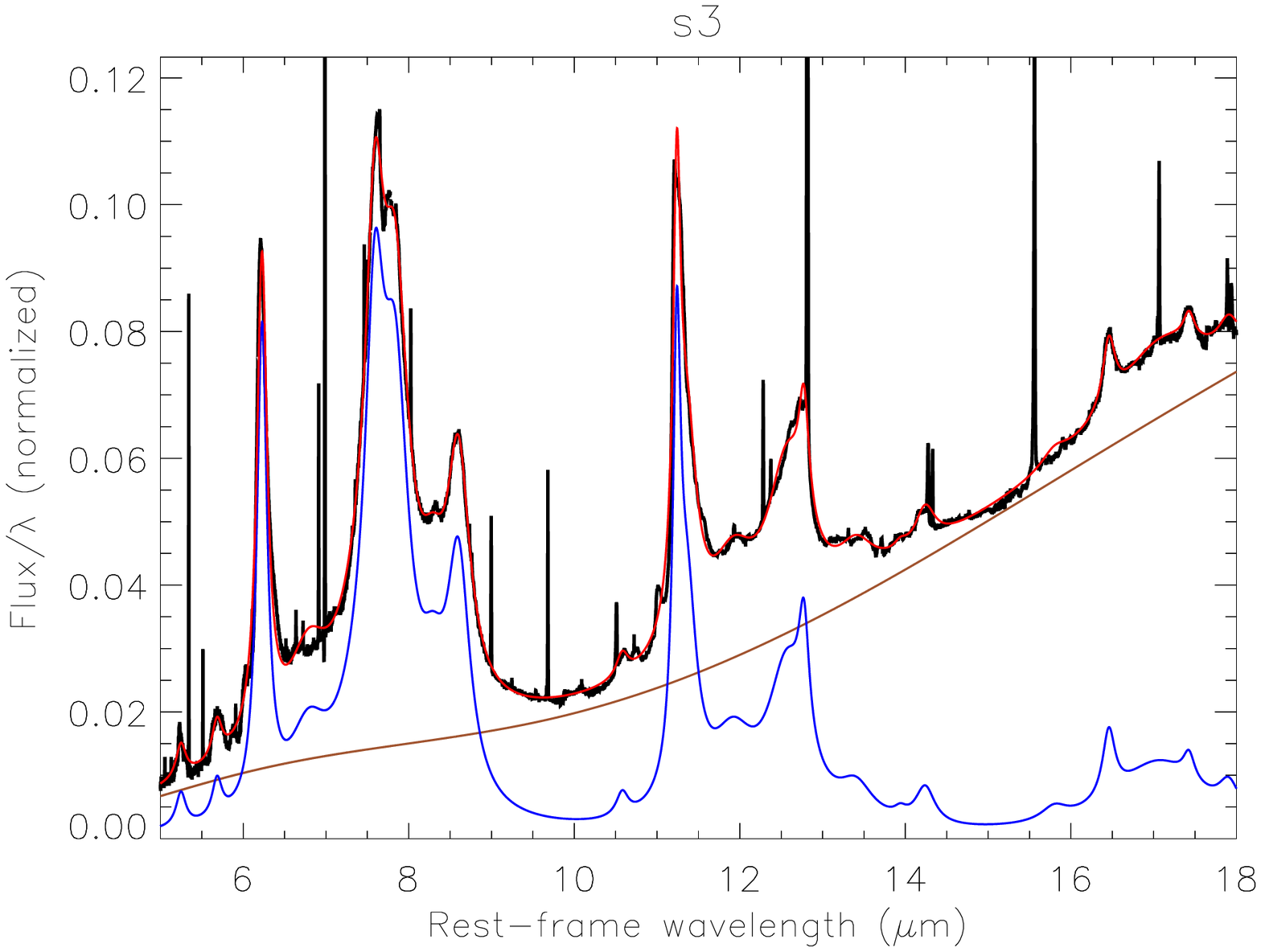}
\includegraphics[width=4.8cm]{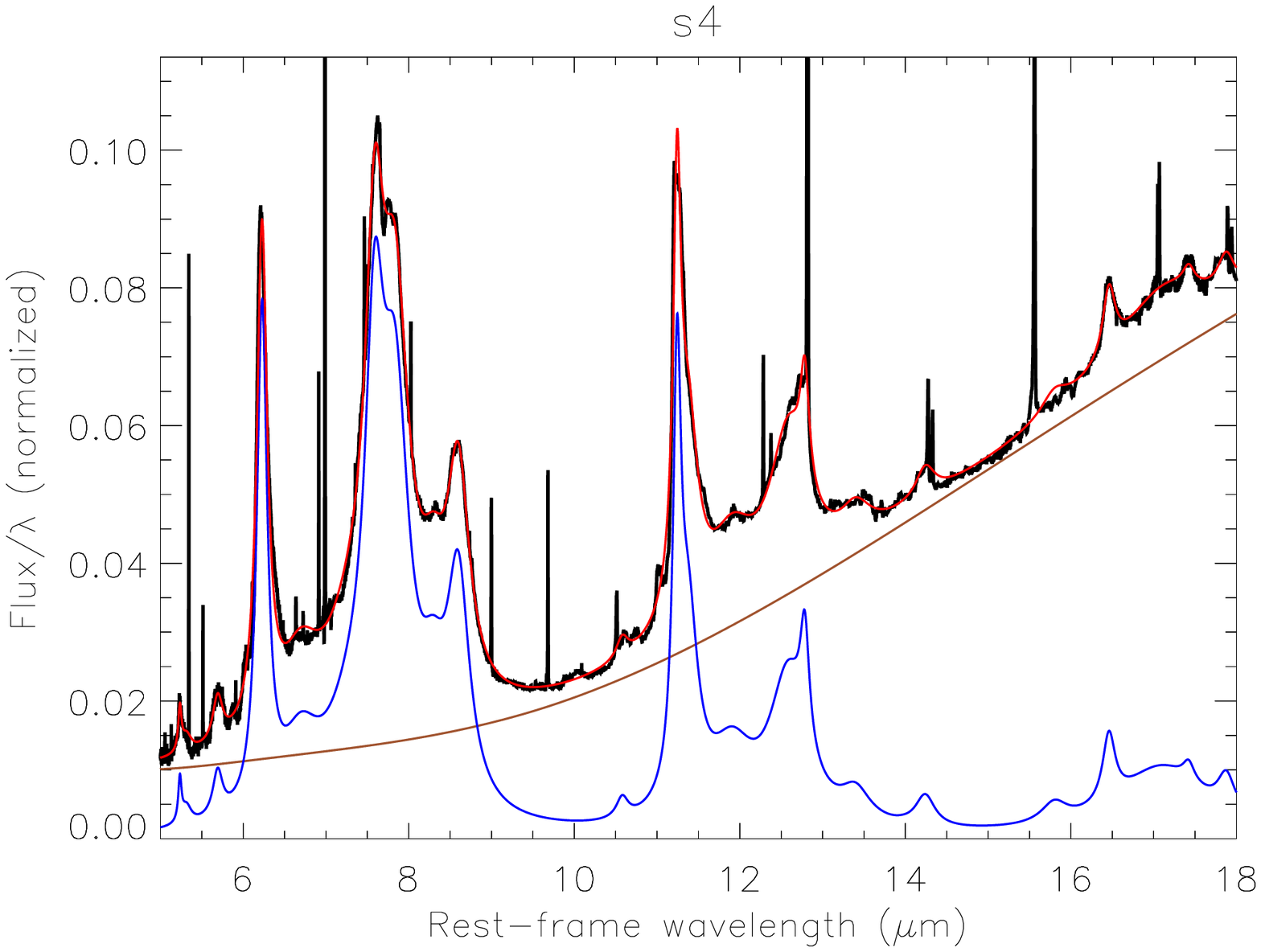}
\includegraphics[width=4.8cm]{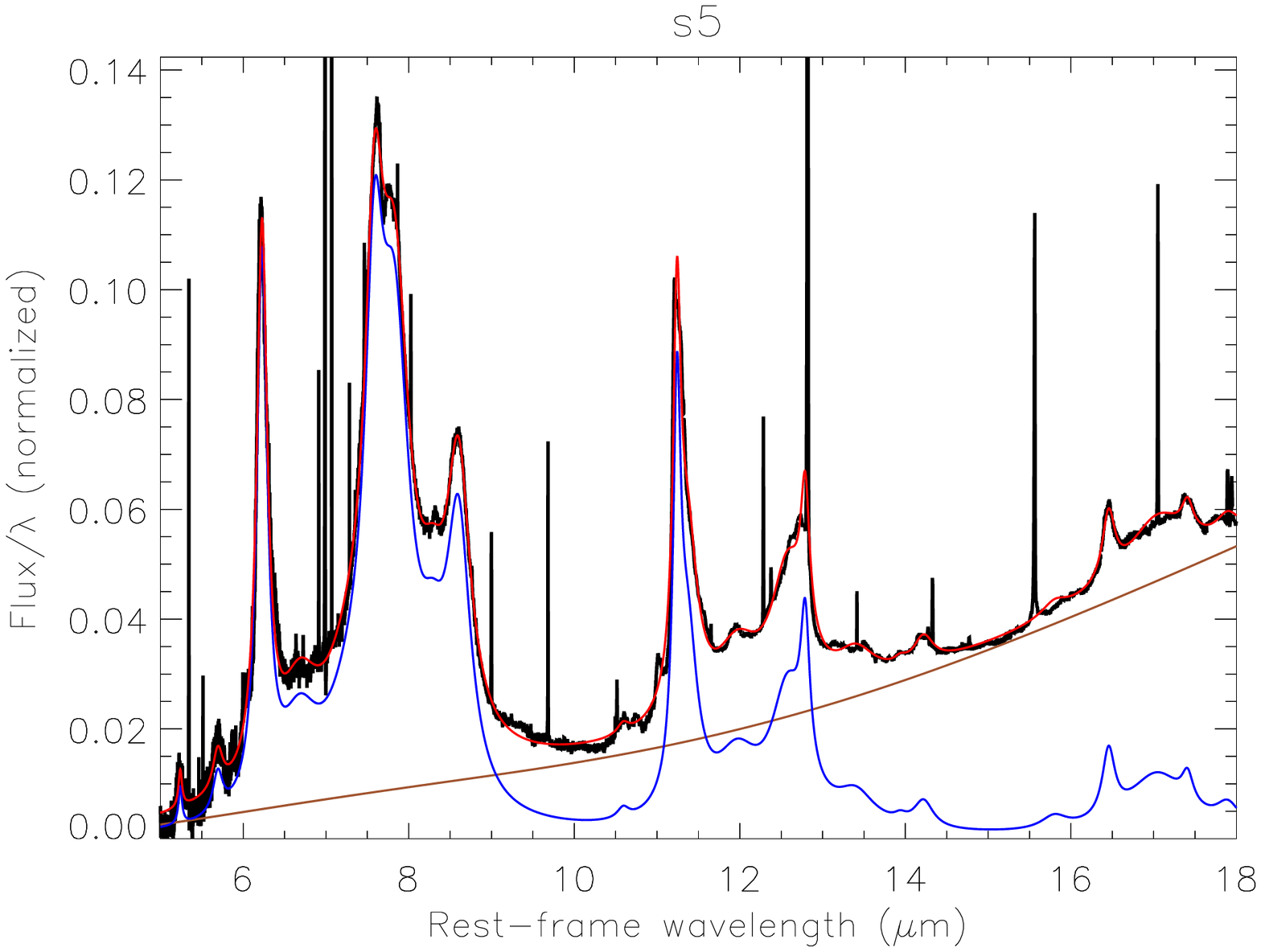}
\includegraphics[width=4.8cm]{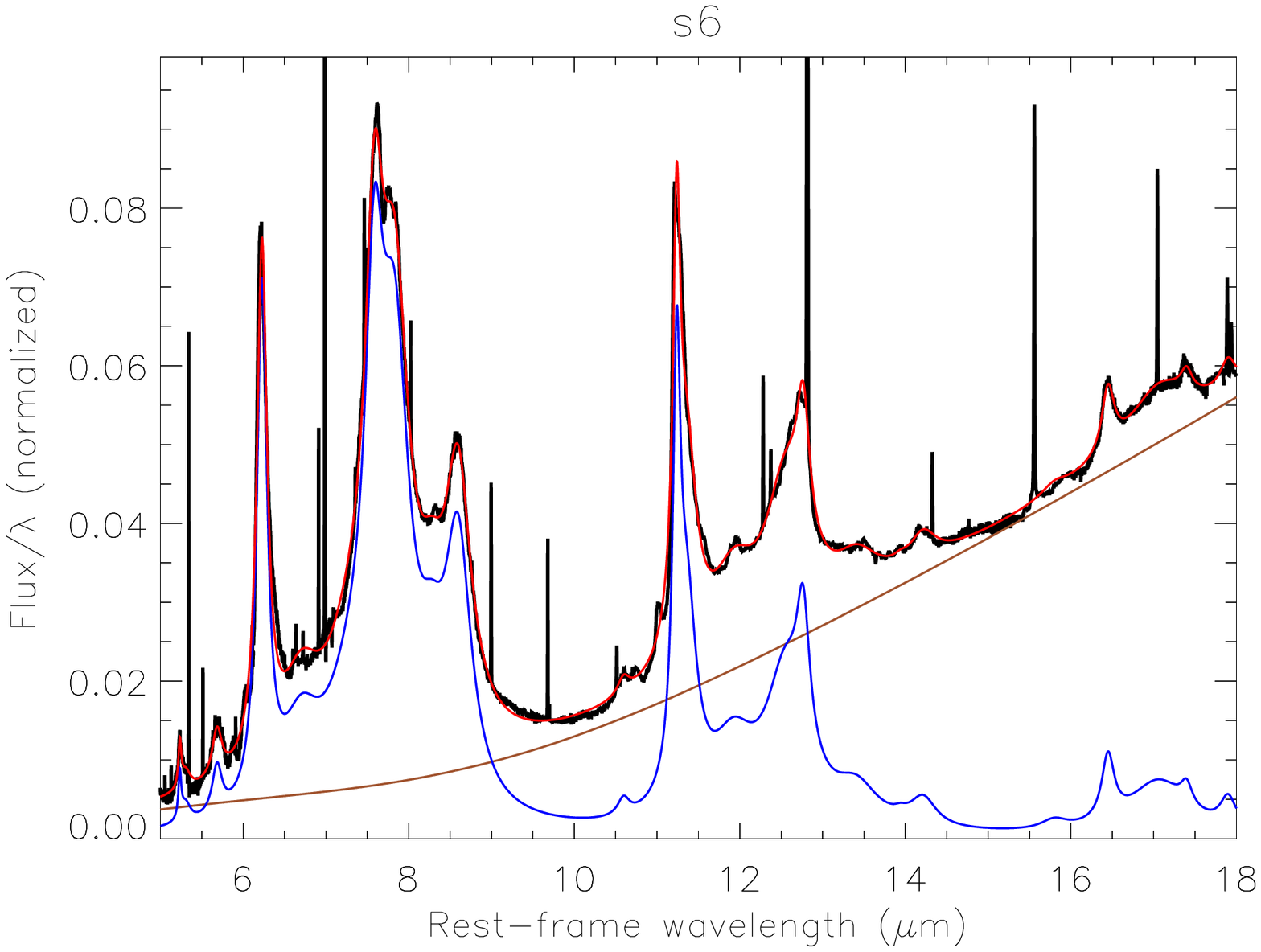}
\includegraphics[width=4.8cm]{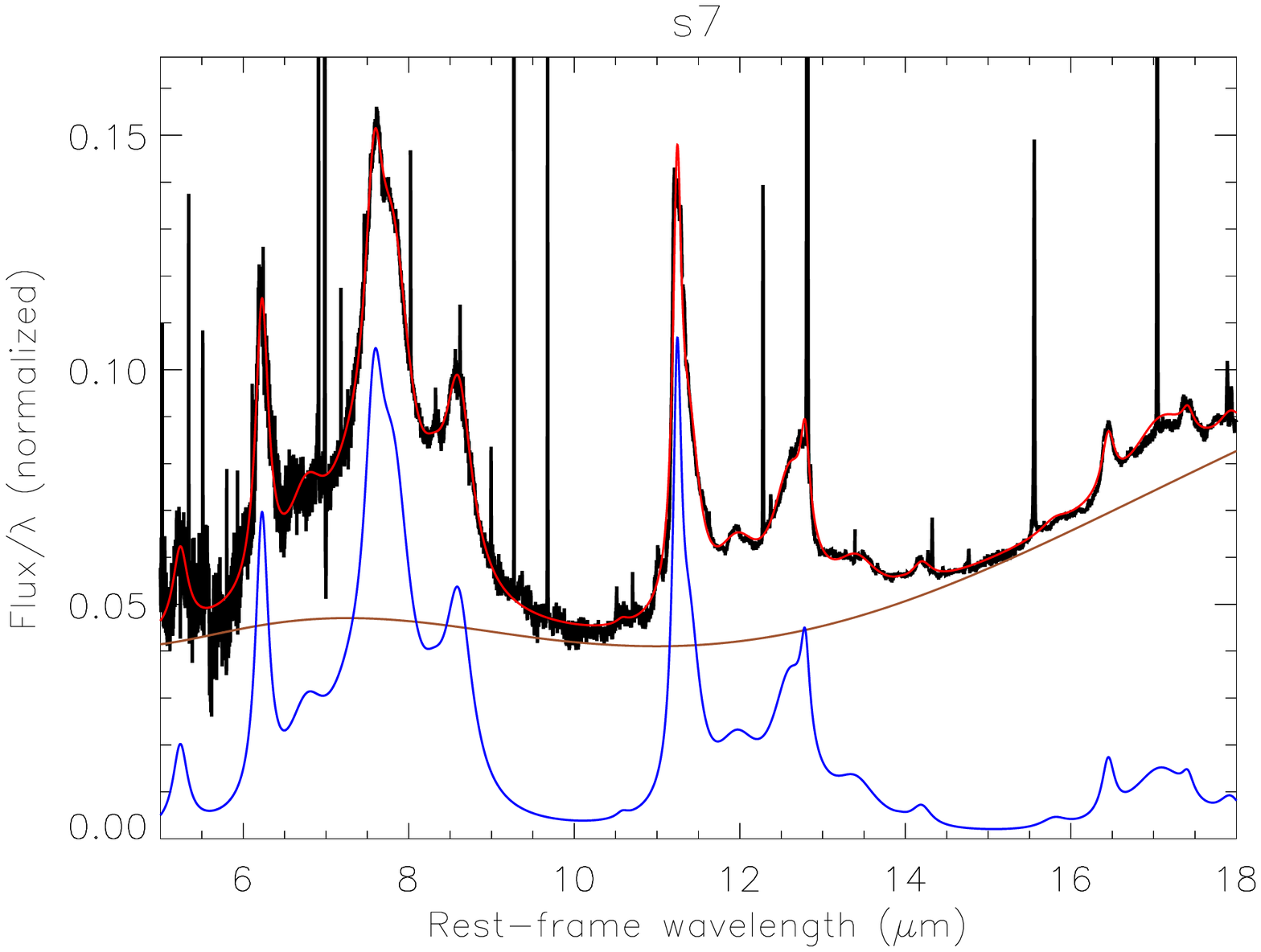}
\includegraphics[width=4.8cm]{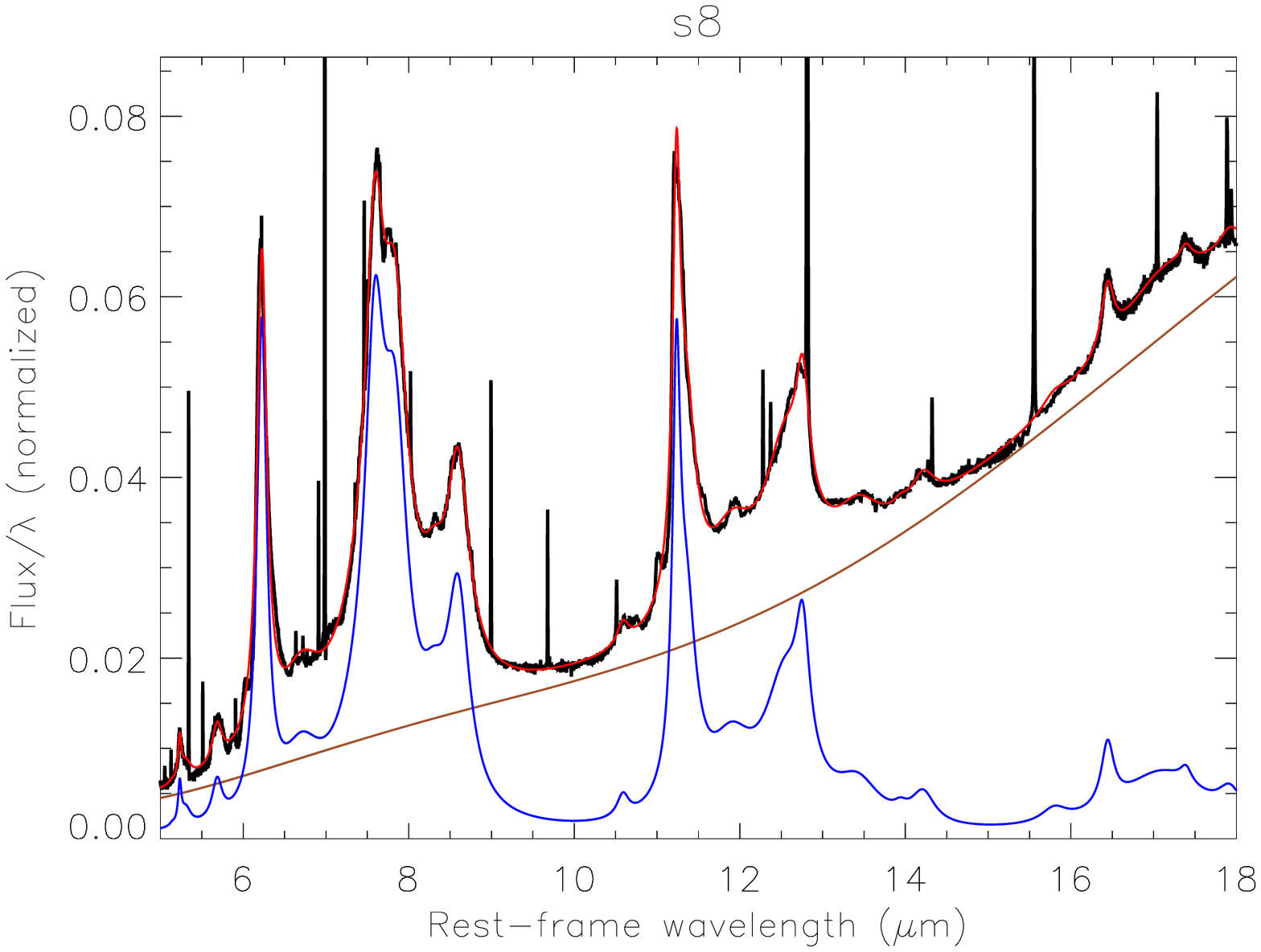}
\includegraphics[width=4.8cm]{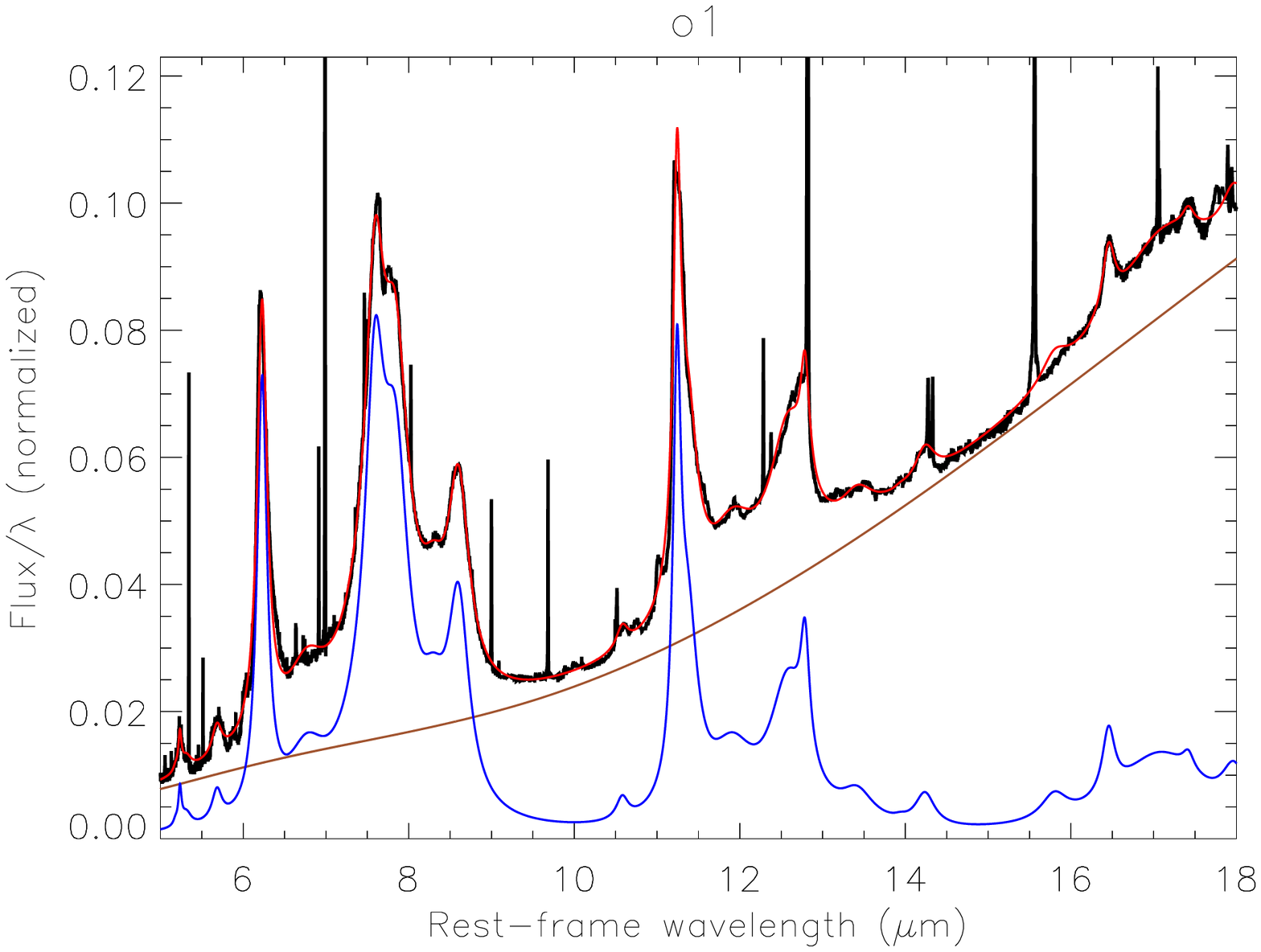}
\includegraphics[width=4.8cm]{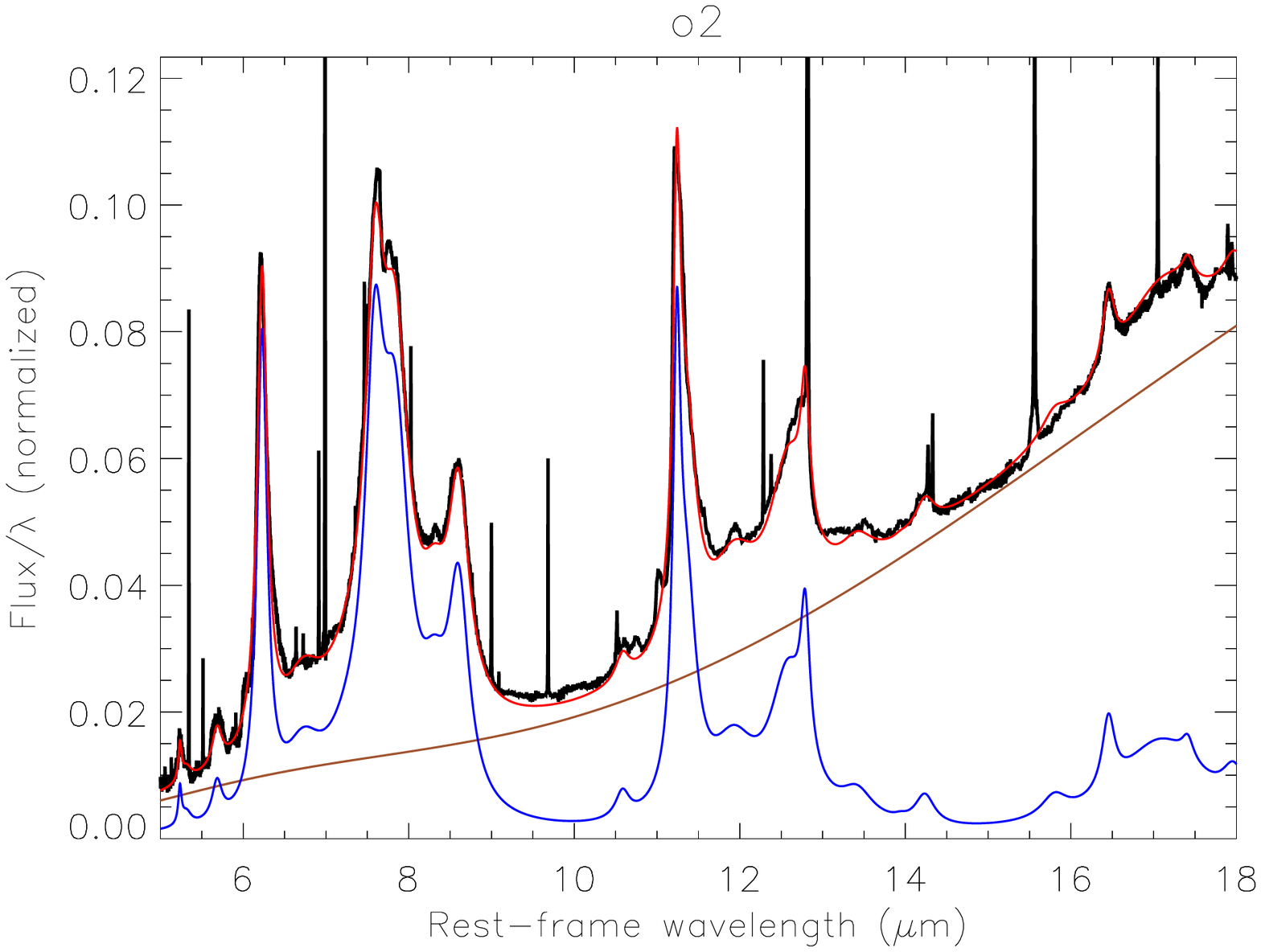}
\includegraphics[width=4.8cm]{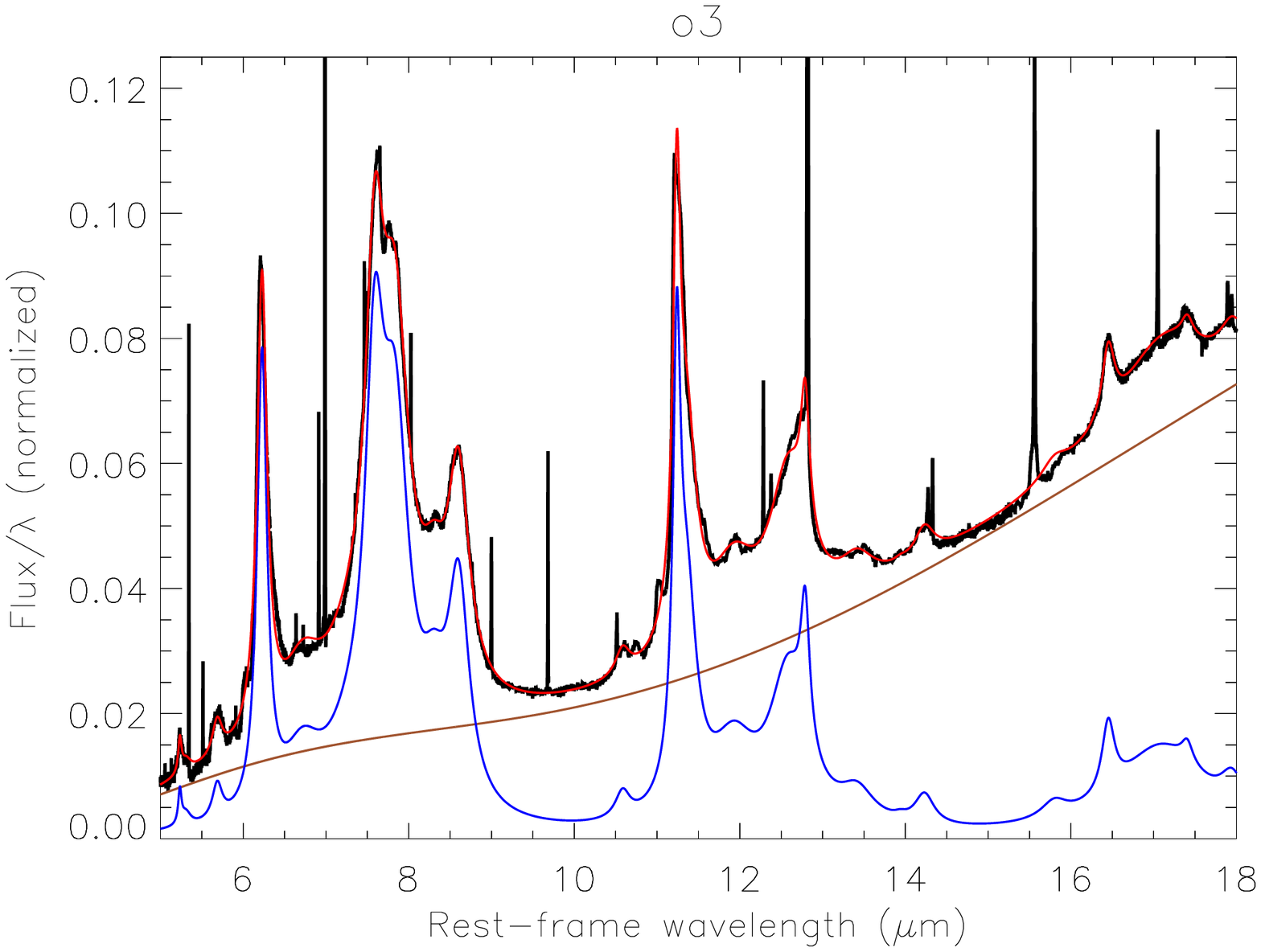}
\includegraphics[width=4.8cm]{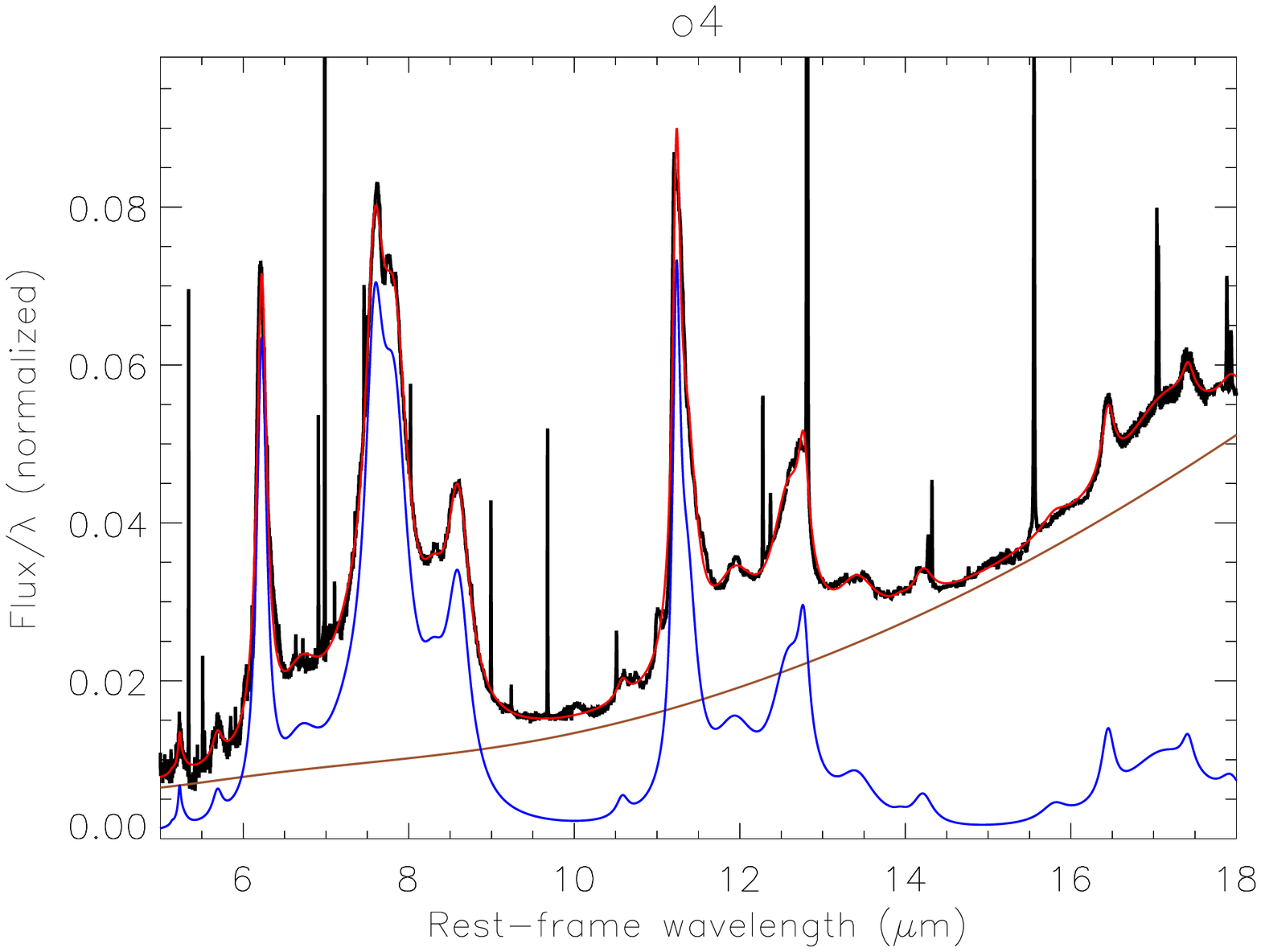}
\includegraphics[width=4.8cm]{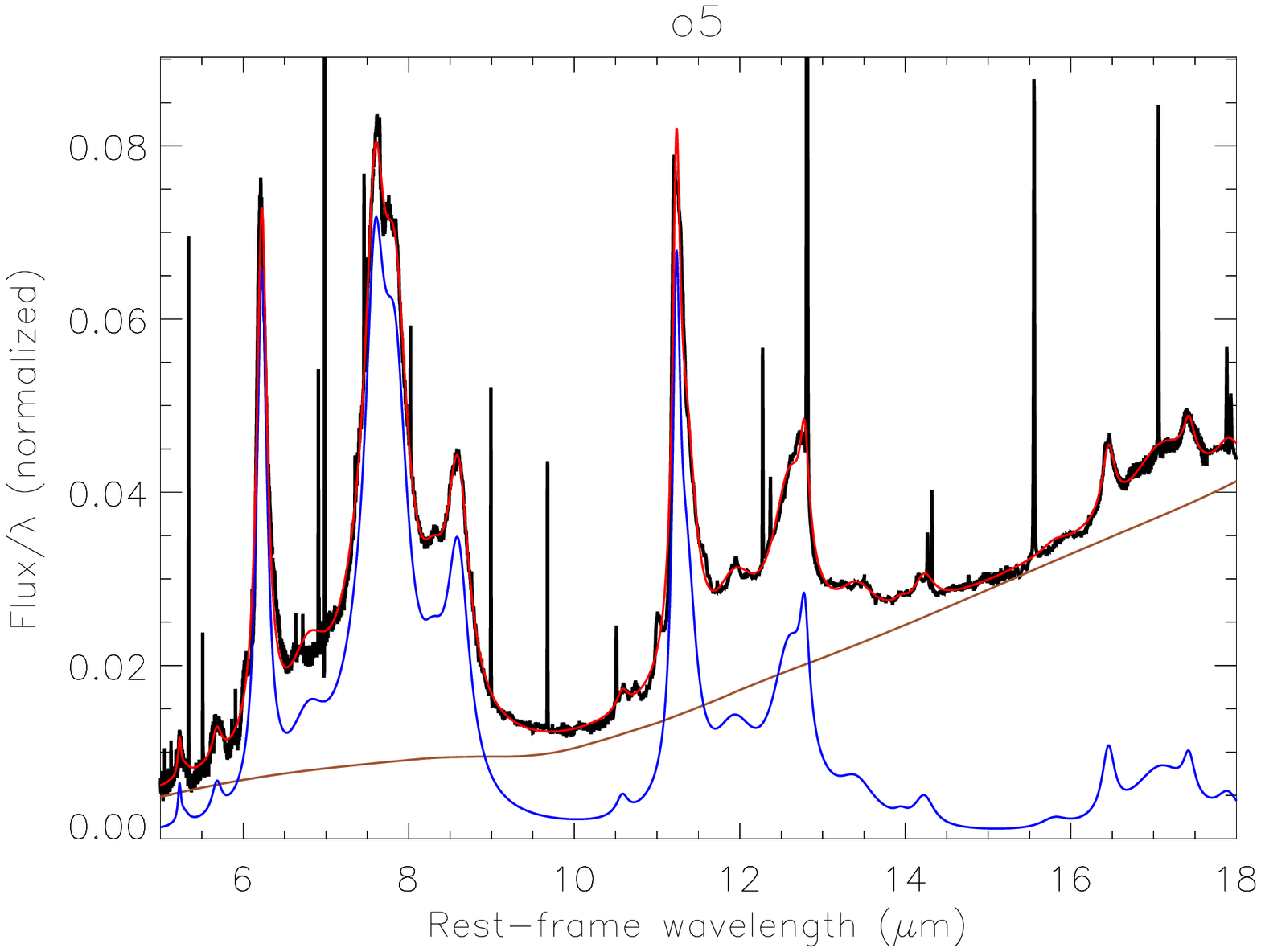}
\includegraphics[width=4.8cm]{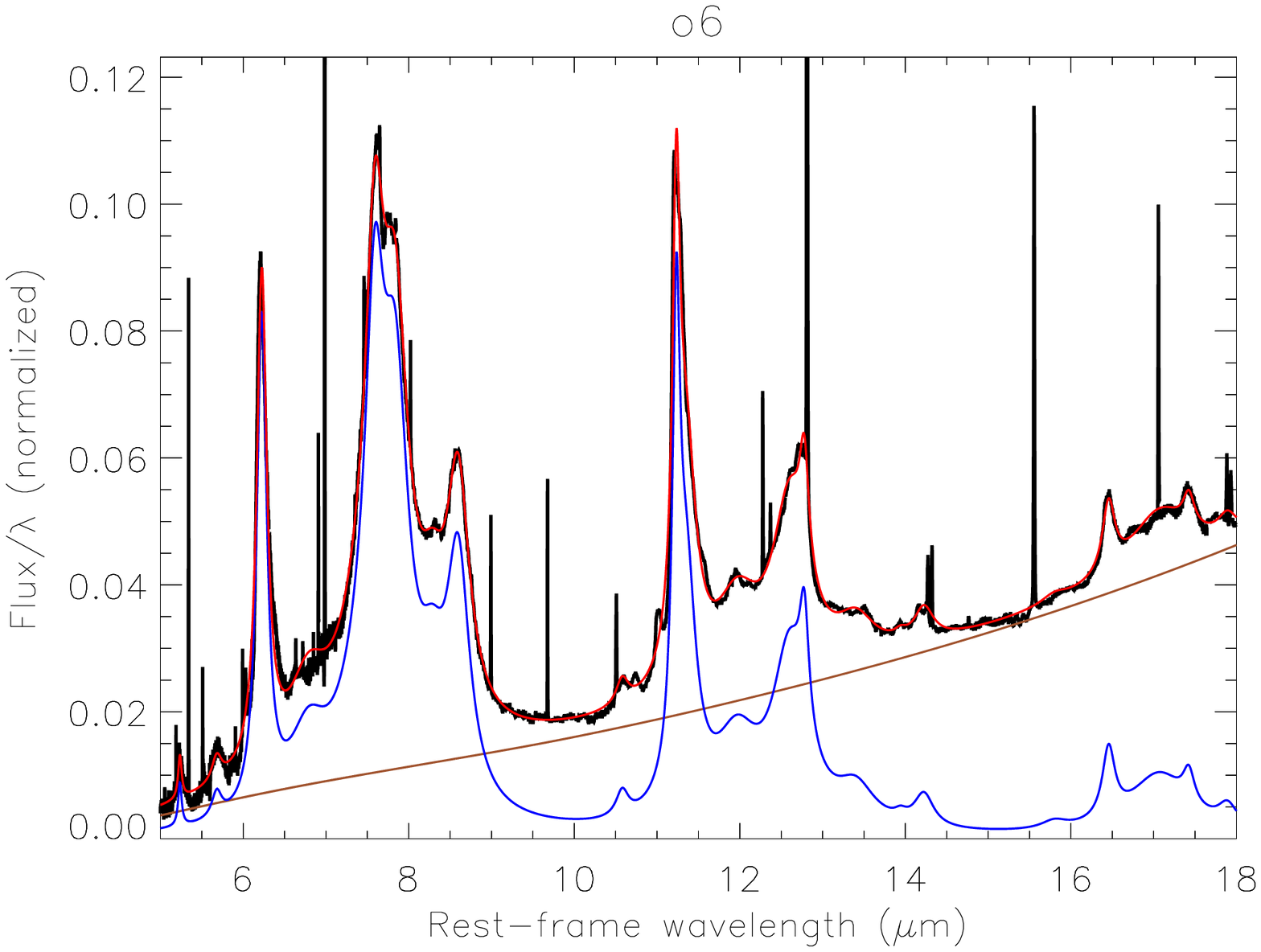}
\par}
\caption{Mid-IR spectral modelling of the circumnuclear regions extracted in this work. The JWST/MRS rest-frame spectra and model fits correspond to the solid black and red lines, respectively. We show the
continuum (solid brown lines) and the fitted PAH features (solid blue lines).}
\label{circumnuclear_fit}
\end{figure*}

In Figs. \ref{nuclear_fit} and \ref{circumnuclear_fit} we show the mid-IR spectral modelling using the modified version of \textsc{PAHFIT}. We note that for the AGN-dominated mid-IR spectrum of NGC\,7319 we used a different approach since its PAH emission is very weak. In particular, we followed the same method as in \citet{Roche07}. The  continuum was fitted with a power law with spectral index S$_\nu\sim\nu^{1.55}$ together with weak silicate emission. 
We used silicate grain profiles derived from the M supergiant $\mu$ Cep, which is representative of the diffuse Galactic
ISM with an optical depth ($\tau_{9.7\mu m}$) = 0.58, to obtain the best fit. Emission from PAH bands, particularly in the 11.3 and 12.7$\mu$m features,  was required for the best fit (Fig. \ref{nuclear_fit2}). In addition, weak emission from the 7.7$\mu$m PAH complex is likely also present, but uncertainties in the shape and level of the underlying continuum preclude quantitative estimates. The wavelengths around obvious narrow emission lines were excluded from the fits.

\begin{figure*}
\centering
\par{

\includegraphics[width=18.8cm]{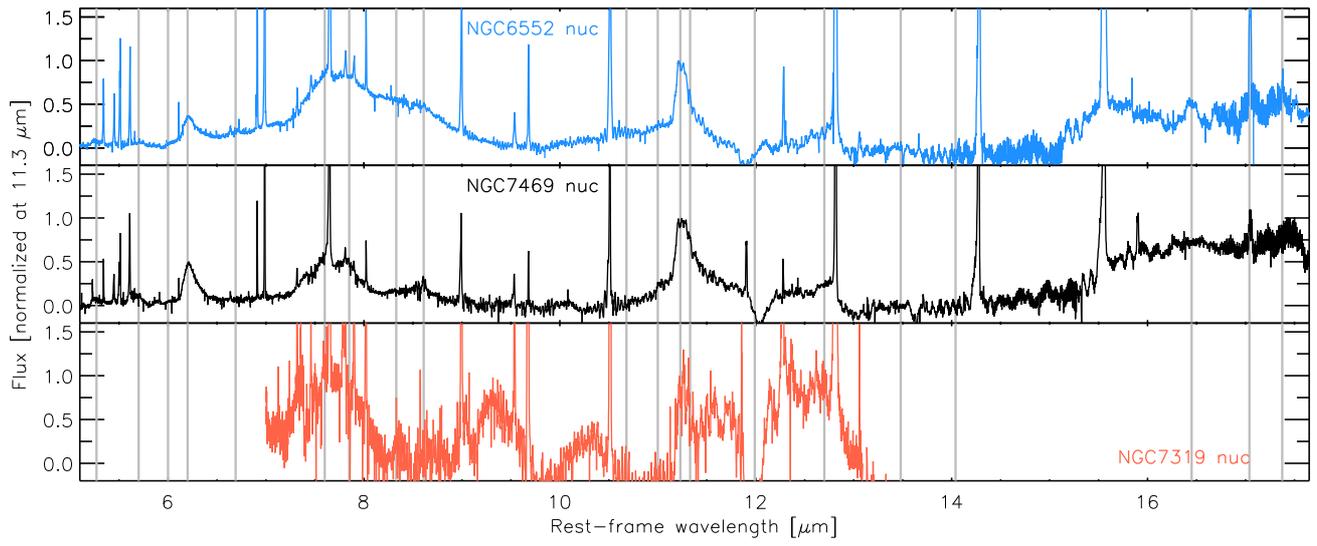}
\par}
\caption{Same as Fig. \ref{pah_profiles} but including the nuclear PAH spectra of NGC\,7319. The red line corresponds to the continuum-subtracted spectra.}
\label{nuclear_residuals}
\end{figure*}

\end{appendix}

\end{document}